\documentclass[smallextended]{svjour3}       % onecolumn (second format)

\smartqed 
\usepackage{booktabs}
\usepackage{color}
\usepackage{framed}
\usepackage{graphicx}
\usepackage{multicol}
\usepackage{appendix}
\usepackage{multirow}
\usepackage{soul}
\usepackage[hyphens]{url}
\usepackage[most]{tcolorbox}
\usepackage{xcolor,colortbl}
\usepackage{listings}
\usepackage{tabularx}
\usepackage[sort,square,comma,numbers]{natbib}

\usepackage{subcaption}

\newcommand\RQOne{What are the characteristics of Helm Charts that are affected by fixable CVEs?}

\newcommand\RQTwo{What strategies do Helm Charts maintainers employ to mitigate fixable CVEs?}

\definecolor{summary_box}{RGB}{128,128,128}
\newtcbtheorem{Summary}{\bfseries Summary of Research Question}{enhanced,drop shadow={black!5!white},
  coltitle=white,
  top=0.3in,
  attach boxed title to top left=
  {xshift=1.5em,yshift=-\tcboxedtitleheight/2},
  boxed title style={size=small,colback=summary_box}
}{summary}

\usepackage{makecell}

\journalname{Empirical Software Engineering}

\begin{document}

\title{Why Not Mitigate Vulnerabilities in Helm Charts?}

\author{Yihao Chen \and Jiahuei Lin \and Bram Adams \and Ahmed E. Hassan}

\institute{Yihao Chen \at 
    School of Computing, Queen's University, Kingston, ON, Canada\\
    \email{yihao.chen@queensu.ca}
    \and
    Jiahuei Lin \at
    School of Computing, Queen's University, Kingston, ON, Canada\\
    \email{jiahuei.lin@queensu.ca}
    \and
    Bram Adams \at School of Computing, Queen's University, Kingston, ON, Canada\\
    \email{bram.adams@queensu.ca}
    \and
    Ahmed E. Hassan \at School of Computing, Queen's University, Kingston, ON, Canada\\
    \email{hassan@queensu.ca}
}

\date{Received: date / Accepted: date}

\maketitle
\begin{abstract}
\textit{Context:} Containerization, a widely used paradigm in modern software development, ensures the resilience of distributed applications by leveraging platforms like Kubernetes for container orchestration. Helm is a package manager for defining, installing, and upgrading Kubernetes applications. A Helm package, namely ``\textit{Chart}'', is a set of pre-configured resources (e.g., network, service interaction) that one could quickly deploy a complex application. However, Helm broadens the attack surface of the distributed applications. 
\textit{Objective:} This study aims to investigate the prevalence of fixable vulnerabilities, the factors related to the vulnerabilities, and current mitigation strategies in Helm Charts.
\textit{Method:} We conduct a mixed-methods investigation on 11,035 Helm Charts affected by 10,982 fixable vulnerabilities. We analyze the complexity of Charts and compare the distribution of vulnerabilities between official and unofficial Charts. Subsequently, we investigate vulnerability mitigation strategies from commits, discussions, issues, pull requests, and automation in the Chart-associated repositories by a grounded theory. 
\textit{Results:} Our findings highlight that the complexity of a Chart correlates with the number of vulnerabilities, and the official Charts do not contain fewer vulnerabilities compared to unofficial Charts. The 10,982 fixable vulnerabilities are at a median of high severity and can be easily exploited. In addition, we identify 11 vulnerability mitigation strategies (e.g., bumping up a version based on a vulnerability report) in three categories, i.e., ad-hoc, automated and informative. Our results indicate that Chart maintainers employ more ad-hoc strategies (6) than those in the other two categories (i.e., 2 automated ones and 3 informative ones). Due to the complexity of a Chart, maintainers are required to investigate where a vulnerability impacts and how to mitigate it. The use of automated strategies is low as automation has limited capability (e.g., a higher number of false positives) in such complex Charts.
\textit{Conclusion:} There is a need for practical automation tools that assist maintainers in mitigating vulnerabilities to reduce manual effort. In addition, Chart maintainers lack incentives to mitigate vulnerabilities, given the complexity of a Chart and a lack of guidelines for mitigation responsibilities. Adopting a shared responsibility model on vulnerability mitigation in the Helm Chart ecosystem would increase its security.

\keywords{Vulnerability \and Security \and Mixed-methods empirical study \and Kubernetes \and Software artifacts}

\end{abstract}

\section{Introduction}
\label{sec:introduction}

Nowadays, software applications are required to be updated and deployed at a fast pace to ensure productive features and better user experience. As such, containerization is revolutionizing the operational landscape of software development and deployment. Containers are lightweight and portable units encapsulating a software application along with its dependencies, libraries, and runtime environments. Such containers enable practitioners to quickly deploy numerous complex and heterogeneous applications and construct artifacts~\cite{burns2016borg}. Datadog, Inc.\footnote{\url{https://www.datadoghq.com/}}, a monitoring and security platform for cloud applications, reported that tens of thousands of its customers ran more than 1.5 billion containers in Nov, 2022.\footnote{\url{https://www.datadoghq.com/about/latest-news/press-releases/datadogs-2022-container-report-finds-organizations-expanding-container-adoption-with-improved-ability-to-scale-and-manage-complex-environments/}}

Kubernetes\footnote{\url{https://kubernetes.io/}} is one of the most popular open-source systems that automate deployment, scaling, and management of containerized applications, i.e., container orchestration platform~\cite{burns2022kubernetes}. Kubernetes orchestrates containers, manages network communication protocols and policies, and monitors service healthiness and resource allocation. Given the richness of containerized applications and the complexity of their runtime environments and interactions, configuring and managing large Kubernetes deployments is challenging. The challenges have led to adoption concerns when applying Kubernetes at a large scale, such as managing software configurations and networking rules~\cite{chen2023practitioners}.

To address such concerns, Helm\footnote{\url{https://helm.sh/}}, a package manager, is designed to manage the deployment of Kubernetes applications~\cite{zerouali2022helm}. Helm uses \textit{Charts}\footnote{\url{https://helm.sh/docs/topics/Charts/}} as reusable packages to encapsulate all necessary Kubernetes resources, including services, deployments, and configurations into a single and reusable bundle~\cite{gokhale2021creating}. Chart users define, install, and manage Kubernetes applications by applying these Helm Charts to the Kubernetes clusters. These Helm Charts could be published and shared on Artifact Hub as a reusable workflow\footnote{\url{https://artifacthub.io/}}, an online repository that hosts artifacts related to Kubernetes~\cite{zerouali2022helm}.

Modern distributed systems, such as service mesh, with their intricate orchestration of microservices, particularly benefit from Helm's streamlined deployment approach. In the 2022 annual Cloud Native Computing Foundation's (CNCF) survey\footnote{\url{https://www.cncf.io/reports/cncf-annual-survey-2022}}, over 90\% of the responded companies and organizations already use or attempt to use Kubernetes, while over 42\% uses Helm to manage deployments. The survey also pointed out that over 40\% of the heavy Kubernetes users (i.e., organizations that orchestrate most of the workloads on Kubernetes) are most concerned about security.

The ease of deployment by Helm Charts broadens the attack surfaces of distributed systems on Kubernetes~\cite{mohallel2016experimenting}. By prepackaging intricate systems into one-click infrastructure deployments, malicious actors could exploit disclosed vulnerabilities that remain unfixed and infiltrate systems deployed via Helm Charts. For example, CVE-2017-5929 (i.e., a critical vulnerability in Logback) in Helm Chart ``JaegerTracing'' remains unfixed (as of Dec 2023, the time of writing this work), even though the patch was released in 2017 and the upstream maintainers suggested an immediate update\footnote{\url{https://logback.qos.ch/news.html#1.2.0}}.

While a large number of research studies has explored the area of vulnerability management, such as vulnerability lifecycles~\cite{anderson2002security, wang2019detecting}, patch development~\cite{zaman2011security, huang2016talos, ozment2006milk} and security practices in software ecosystems (e.g., NPM~\cite{zahan2022weak, zimmermann2019small}, GitHub~\cite{acar2017security}, Linux distributions \cite{lin2023vulnerability}), a few research efforts investigated vulnerabilities in containerization~\cite{zerouali2022helm, gokhale2021creating, wist2021vulnerability}. A previous study~\cite{gokhale2021creating} suggested that Helm Charts are loosely assembled by numerous dependencies maintained by different parties. \citet{wist2021vulnerability} reported that official images on the Docker Hub are more secure than unofficial ones regarding the number of high-to-critical severity vulnerabilities. While \citet{zerouali2022helm} examined the security status of overall Helm Charts regarding the number of vulnerabilities, the authors did not investigate why the vulnerabilities are not yet mitigated and the reasons behind that, even though a fix is available. Given that Helm Charts serve as the final software artifact before large-scale deployments, the unique nuances of the Helm ecosystem warrant dedicated investigation.

In this study, we employ a mixed-method approach to address this gap. We perform an in-depth quantitative analysis on 11,035 Helm Charts in Artifact Hub affected by at least 13,095 unique vulnerabilities (see Table \ref{table:chart-report-metrics}). Then, we perform a qualitative Grounded Theory (GT) study from diverse data types (e.g., commits, pull requsts, automation) in 90 GitHub repositories associated with 686 Helm Charts. We will elicit the GT methodology (i.e., data sampling, coding phases and saturation) step by step in Section~\ref{sec:methodology}. We investigate the prevalence of vulnerabilities, their correlated factors (e.g., severity), and probe vulnerability mitigation practices via a GT methodology that is widely used in previous empirical studies to understand complex phenomenon~\cite{hoda2012developing,carver2007use,diaz2023applying,stol2016grounded,adolph2011using,zimmermann2023grounded,foundjem2023grounded}. The GT approach complements our quantitative analysis by eliciting practitioner perspectives (Chart maintainers and Chart users). 

Our key findings and their implications include the following: 
\begin{itemize}
\item We observe a high prevalence of vulnerabilities, with 13,095 unique CVEs (10,982, 83.9\% are fixable) identified across the 6,202 (56\%) Helm Charts with a security report. The high number of CVEs indicates widespread vulnerabilities across the ecosystem. Notably, although 56\% of the Charts utilize security reports, only 80 (1\%) of these Charts use security fix indicators (e.g., writing security fixes in changelogs).

\item The distribution of CVEs per Helm Chart is skewed (i.e., the skewness coefficient of 30.375). Interestingly, the number of CVEs in a Chart significantly correlates with the number of dependent packages within all the images in a Chart (partial correlation coefficient of 0.85, $p\text{-value}<0.001$), but does not correlate with the number of container images (partial correlation of -0.47, $p\text{-value}<0.001$). Such a correlation suggests that practitioners should take into account the number of dependent packages rather than the number of images.

\item The statistical analysis shows significant differences in fixable CVE severity distributions across officiality (p$<$0.001). However, the effect sizes are negligible based on Cliff's Delta, indicating that official Charts are not less vulnerable than unofficial ones.

\item Our qualitative analysis identifies 11 mitigation strategies grouped by 3 categories from the 90 Chart maintenance repositories (that are associated with 686 Helm Charts): \textbf{Ad-hoc} (6 strategies), \textbf{Automated} (2 strategies), and \textbf{Informative} (3 strategies). Chart maintainers generally rely on manual effort rather than automation to mitigate vulnerabilities. Given the high complexity of a Chart, maintainers may lack expertise related to security and wait for a fix from the upstream dependency. Adopting a shared responsibility model would reduce the overall maintenance effort across maintainers and improve the Chart ecosystem's security.

\item The low usage of automated strategies (i.e., 2 strategies: always release the latest dependencies and utilize CVE scanners) highlights the limited capability of existing automated tools and the limited choice of automation. Chart maintainers need to make additional efforts to fix the issues introduced by the automation tools (e.g., false positives or incompatibilities). Thus, Chart maintainers prioritize stability and feature requirements over security. 

\item We provide the replication package of our study with the main study artifacts, including the data collection scripts, data analysis scripts and spreadsheets for the grounded theory process~\cite{chen_lin_adams_hassan_2023}.

\end{itemize}

We present the remainder of this paper as follows. Section~\ref{sec:background} introduces the background of the Helm Chart ecosystem and CVEs. Section~\ref{sec:related-work} discusses prior related works to our study. Section~\ref{sec:methodology} motivates our study with two research questions to be answered progressively. We also follow each detailed step with analysis protocols, data collection techniques and corresponding artifacts. Sections~\ref{sec:rq1-results} to~\ref{sec:rq3-results} discuss the evaluation and results discussions for each research question. Section~\ref{sec:implications} summarizes the key findings and further gives practical and actionable suggestions to researchers in the domain and practitioners who are looking to use/maintain Helm Charts with better security practices in mind. Section~\ref{sec:threats} acknowledges and presents the potential threats to our study and justifies the methodologies we adopted to best complement some of the shortcomings. Finally, Section~\ref{sec:conclusions} concludes our findings and looks into potential research opportunities in the future.
\section{Background}
\label{sec:background}

\subsection{Helm Charts, Chart Dependencies, and Vulnerability Statuses}
\label{subsec:background-charts-dependencies}

A Helm Chart is composed of a collection of \textit{.yaml} files that describe a set of required Kubernetes resources (e.g., predefined configurations and their dependencies) for deployments\footnote{\url{https://helm.sh/docs/topics/charts/#the-chart-file-structure}}. A Chart consists of metadata (e.g., version, the home page URL), a set of required packages (e.g., NGINX, MySQL) and/or required Charts. We refer to the required packages as ``\textit{package dependencies}'' and the required Charts as ``\textit{Chart dependencies}''. Due to the nested structure in a Chart, the complexity of a Helm Chart varies across applications. Some Helm Charts are simple, such as encapsulating an application containing a single Chart in one file, while others encompass an entire application stack composed of many interconnected Charts. These Charts have their own packages and Chart dependencies that several parties of maintainers maintain, each with a different release schedule. As the number of dependencies (i.e., both Chart and package dependencies) in a Helm Chart increases, the likelihood of the Helm Chart being affected by vulnerabilities increases, resulting in a higher possibility of getting attacks~\cite{mohallel2016experimenting}. Note that in the rest of this paper, we refer to Helm Charts as ``\textit{Charts}".

\begin{table}[h!]
\centering
\caption{Three statuses of a vulnerability in a Helm Chart along with the number of such status of vulnerabilities in our dataset. Note that this study focuses on the \textbf{fixable} vulnerabilities.}
\label{table:unfixed_define}
\begin{tabular}{lp{6.8cm}r}
\toprule
 \textbf{Status} & \textbf{Description} & \textbf{Count}\\ 
\midrule
Fixed & A fixed vulnerability indicates that a Chart was vulnerable, but a fix has been incorporated.  & NA \\ 
\textbf{Fixable} & A fixable vulnerability indicates that a Chart is affected by the vulnerability when a vulnerability fix has been developed, but the Chart maintainers have not yet incorporated the fix. &  10,982 (83.9\%)\\
Not-patch-yet & A not-patch-yet vulnerability represents the vulnerability has not yet been developed. & 2,873 (16.1\%) \\
\midrule
Sum & & 13,095\\
\bottomrule
\end{tabular}
\end{table}

Table~\ref{table:unfixed_define} defines three statuses of a vulnerability in Charts, i.e., \textit{fixed}, \textit{fixable} and \textit{not-patch-yet}. A \textit{fixed} vulnerability in a Chart indicates that the vulnerability mitigation has been applied. A \textit{fixable} vulnerability represents the vulnerable upstream dependency that has released a fix that the Chart maintainers could mitigate the vulnerability. However, the maintainers were not able to incorporate the fix due to certain reasons (e.g., resource constraints). A \textit{not-patch-yet} vulnerability indicates that the fix has not yet been developed. Note that we refer to \textit{unfixed} vulnerabilities as the union set of \textit{fixable} and \textit{not-patch-yet} vulnerabilities.

In this study, we are interested in the \textit{fixable} vulnerabilities that remain unfixed. If they remain unfixed, such fixable vulnerabilities could pose a significant risk to deployed systems~\cite{zerouali2019impact}. It is crucial to understand fixable vulnerabilities in Helm Charts, so practitioners could derive better vulnerability management strategies and researchers could investigate practices to improve the security of Charts.

\subsection{Artifact Hub, GitHub and Cloud Native Computing Foundation (CNCF)}

Artifact Hub\footnote{\url{https://artifacthub.io/}} is a public repository where one can search for reusable cloud-native artifacts, like Helm Charts, install them and publish their own artifacts. Artifact Hub has more than 11,035 Charts available to users, with each of these Charts having multiple versions (median of 4 versions). 

Like traditional software artifacts, Charts require new releases to address bugs and ship new features. The maintenance process includes monitoring vulnerabilities and their fixes, upgrading the changes made by upstream dependencies, incorporating bug fixes, and shipping new features. As this process is complex, Chart maintainers often use GitHub to host and maintain their Charts. In addition, Helm offers command-line tools to ease the maintenance of Charts, as such tools could be seamlessly integrated into the maintenance process (e.g., GitHub Actions) on GitHub. Therefore, GitHub becomes a richer source of information that encodes maintenance activities (e.g., CVE mitigation strategies) and the rationale for employing a strategy. Therefore, in this study, we collect Helm Charts from Artifact Hub and the associated maintenance repositories from GitHub.

CNCF\footnote{\url{https://www.cncf.io/}} is a non-profit organization that was founded to facilitate the development and deployment of cloud-native technologies that empower companies to run scalable applications on public, private, and/or hybrid cloud. Containerization is one of the cloud-native technologies. The CNCF has hosted 20 graduated projects, 35 incubating and 102 sandbox projects\footnote{\url{https://www.cncf.io/reports/cncf-annual-report-2022/#}}. CNCF defines criteria to evaluate the maturity of a project that is considered stable and used successfully in production environments, i.e., graduated projects.\footnote{\url{https://www.cncf.io/projects/} } Kubernetes is one of the graduated projects now maintained by CNCF.
Artifact Hub is a sandbox project that was founded in June 2020.\footnote{\url{https://www.cncf.io/projects/artifact-hub/}}

\subsection{Common Vulnerability Scoring System (CVSS) Scoring and CVSS Metrics}
\label{sec:sub_cvss}
The Common Vulnerability Scoring System (CVSS) is a standard used to measure the severity of a vulnerability\footnote{\url{https://nvd.nist.gov/vuln-metrics/cvss}}. The CVSS standard has been widely used in security-related studies~\cite{lin2023vulnerability, shahzad2012large, zhang2021study}. 
We select the CVSS V3 scores as the V3 is the latest standard from the National Vulnerability Database (NVD) that went into effect in 2019~\cite{first2015common}. We only consider the base scores since the temporal and environmental scores that differ within each application are beyond reasonable effort to collect, given the large number of Charts (11,035 Helm Charts) in the dataset (see Table \ref{table:Chart-report-metrics}).

The CVSS V3 base score is a quantitative representation of the severity of a vulnerability. It provides a numerical score ranging from 0 to 10, with 10 being the most severe. Specifically, the score is calculated through eight categorical metrics, i.e., 4 exploitability metrics, 1 scope metric and 3 impact metrics assigned by security experts to evaluate a unique characteristic of a given vulnerability~\cite{first2015common}. We note the exploitability metrics with a prefix of ``(E)'', the impact metrics with a prefix of ``(I)'', and the scope metric with a prefix of ``(S)'' and detail them below:

\begin{itemize}
    \item (E) Attack Vector: Through which means the vulnerability can be exploited, such as locally or remotely.
    \item (E) Attack Complexity: The complexity of the attack required to exploit the vulnerability.
    \item (E) Privileges Required: The level of privileges required for an attacker to exploit the vulnerability.
    \item (E) User Interaction: Whether user interaction is required to exploit the vulnerability.
    \item (S) Scope: Whether a vulnerability in one component impacts resources beyond its scope.
    \item (I) Confidentiality: The impact on confidentiality of a successfully exploited vulnerability.
    \item (I) Integrity: The impact on system integrity of a successfully exploited vulnerability.
    \item (I) Availability: The impact on the availability of a successfully exploited vulnerability.
\end{itemize}

\section{Related Work}
\label{sec:related-work}

Our research extends the current body of knowledge on Helm Chart vulnerabilities, package manager security research, and supply chain security in Infrastructure as Code (IaC). Here, we summarize key related works within these areas and motivate our study.

\subsection{Vulnerabilities in Helm Charts}

A few studies~\cite{blaise2022stay, zerouali2022helm} have started to uncover the complex landscape of Helm Chart vulnerabilities. \citet{blaise2022stay} introduced a methodology to graph and quantify the attack risks associated with Helm Charts. The authors discovered that Helm Charts can be vulnerable through sophisticated attack paths due to their complexity. Our work further empirically substantiates their findings, revealing a high prevalence of identified vulnerabilities (i.e., fixable CVEs) in Helm Charts. Furthermore, we observe a general reluctance among maintainers to engage in sophisticated vulnerability prevention when producing Helm Charts, even though the vulnerabilities are fixable. Instead, current Chart maintainers depend predominantly on ad-hoc strategies that expose shortcomings. For example, bumping up dependency versions on CVE reports in an ad-hoc way may pose security risks since Helm Charts CVEs are easily exploitable and of moderate impact.

Another empirical study~\cite{zerouali2022helm} investigated the growth, reuse, dependencies, as well as outdatedness of the container images used within them. Their research highlighted that despite the exponential growth of Helm Charts, most lack popularity and open-source licenses. About 90\% of the Charts were found to be exposed to vulnerabilities affecting their images. Our study confirms their findings and further finds that over 80\% of the vulnerabilities are fixable. In addition, through a comprehensive mixed-methods study, with statistical analysis and a qualitative grounded theory approach, our study dives deeper into the reasons behind the persistence of these fixable vulnerabilities and the complex reasons behind the mitigation.

\subsection{Package Manager Security Research}
Extensive studies have been conducted across various package management ecosystems, such as NPM~\cite{zahan2022weak, chinthanet2021lags}, Ruby~\cite{prana2021out}, and Python~\cite{alfadel2023empirical, ruohonen2018empirical}, among others, revealing inherent security challenges. Package managers, due to their reliance on dependencies, can often act as vectors for vulnerabilities. Unlike traditional package managers, Helm Charts package up not only an entire application but also its dependencies, configurations (e.g., the number of serving nodes) and required resources (e.g., databases, network, other applications) for deployments on Kubernetes (as discussed in Section \ref{sec:background}). Such complexity introduces additional attack vectors that increase security risks. Furthermore, Helm Charts are crucial in the software development life cycle, functioning as the last software artifact before deployment. As the Helm Charts ecosystem is loosely organized and different parties maintain the dependencies of a Chart, vulnerability management becomes more challenging. Our study extends this line of research by focusing on these unique aspects of Helm and their implications for vulnerability management.

\subsection{Supply Chain Security in Infrastructure as Code (IaC) systems}
Supply chain security in IaC systems represents another key dimension related to our study. IaC systems enable the definition and provisioning of complex system components through code, bringing new perspectives on the problem of security vulnerabilities. Previous studies\cite{rahman2019seven,lepiller2021analyzing,guerriero2019adoption} have explored inherent challenges of security and compliance in IaC scripts, such as those used in Terraform\footnote{\url{https://www.terraform.io/}} and Chef\footnote{\url{https://www.chef.io/}}. However, Helm Charts as a form of IaC for Kubernetes environments has not well explored. In this study, 
we conduct a novel and comprehensive investigation of vulnerability management in the Helm Chart ecosystem to provide practical takeaways for researchers, Chart maintainers and users.
\section{Study Design}
\label{sec:methodology}

\subsection{Goal and Research Questions}
\label{sec:methodology:subsec:goal}

We rely on the Goal/Question/Metric template~\cite{basili1988tame} to define our study goal as follows: \textit{to analyze} the fixable CVEs in Helm Charts and the mitigation strategies employed by Chart maintainers; \textit{for the purpose of} understanding the prevalence of fixable CVEs in Helm Charts; \textit{with respect to} identifying the characteristics of fixable CVEs through related metrics (e.g., CVSS scores, the number of package dependencies) and informative security features (e.g., changelogs), and available strategies to mitigate such CVEs in the maintenance process; \textit{from the point of view of} Chart maintainers and software engineering researchers; \textit{in the context of} Helm Charts, their fixable vulnerabilities and the maintenance activities of maintainers in the associated GitHub repositories.

We introduce our two research questions (RQs) as follows:

\medskip \noindent \textbf{RQ1: \RQOne}

\smallskip \noindent \textit{Motivation:} As prior work~\cite{shu2017study, wist2021vulnerability} pointed out the growing concerns regarding the prevalence of vulnerabilities in the containerization domain, this RQ explores such concerns in Helm Charts. As Helm has become a pivotal tool in the Kubernetes ecosystem, the security of its Charts is important for both Chart maintainers and users. Understanding the characteristics of the fixable vulnerabilities that remain unfixed in Helm Charts provides insights into the awareness of security risks and the selection of Charts.

\begin{table}[t]
\centering
\caption{Basic statistics about the studied 11,035 Helm Charts affected by at least 13,095 unique vulnerabilities. Note that we cannot collect the number of vulnerabilities in the Charts \textbf{without} a security report.}
\label{table:chart-report-metrics}
\begin{tabular}{p{3cm}rrrrr}
\toprule
Metric & Official$^1$ & Unofficial$^1$ & Sum$^1$$^2$ & CVE Count \\
\midrule
\# Charts \textbf{with} security reports & 150 (65.2\%) & 6,052 (54.8\%) & 6,202 (56\%) & 13,095 \\
\# Charts \textbf{without} security reports & 80 (34.8\%) & 4983 (45.2\%) & 4,833 (44\%) & N/A \\
\midrule
Sum & 230 & 10,805 & 11,035 & N/A &\\
\bottomrule
\end{tabular}

\raggedright $^1$ The percentage is computed relative to the cell in the last row. \\
\raggedright $^2$ The number in a cell indicates the sum of the number of official and unofficial Charts. \\
\label{table:Chart-report-metrics}
\end{table}

\smallskip \noindent \textit{Approach:} We obtain a large dataset of 11,035 Helm Charts (see Table \ref{table:chart-report-metrics}) that are publicly available on Artifact Hub. We extract these Charts and their associated CVEs via the corresponding security reports (6,202 Charts disclose security reports). To evaluate the prevalence of fixable CVEs, we calculate the number of fixable CVEs in each Chart, analyze whether the Charts are affected by common fixable CVEs and whether the complexity (e.g., Chart dependencies and package dependencies) of a Chart is correlated with the number of fixable CVEs. We also use the CVSS score of these fixable CVEs, which are widely used in many prior work~\cite{lin2023vulnerability, joh2011defining, lin2023coordination}, to investigate their severity, exploitability, and impact scopes. As Artifact Hub suggests disclosing vulnerability fixes in changelogs\footnote{\url{https://artifacthub.io/docs/topics/annotations/helm/}} and GitHub provides security-related informative features (e.g., SECURITY.md)\footnote{\url{https://docs.github.com/en/code-security/getting-started/github-security-features}}, we measure the use of such informative security features in Charts.

\medskip \noindent \textbf{RQ2: \RQTwo}

\smallskip \noindent \textit{Motivation:} The results in RQ1 indicate fixable vulnerabilities (83.9\%) are dominant in Helm Charts, and the complexity of nested Chart and package dependencies correlates with the number of vulnerabilities. Given that, to understand how maintainers mitigate such a large number of vulnerabilities in a Chart, we conduct a grounded theory (GT) study~\cite{zimmermann2023grounded, stol2016grounded} to provide qualitative insights into mitigation strategies. Providing such insights could improve current security practices in the containerization domain and be shared across Chart maintainers.

\smallskip \noindent \textit{Approach:} Prior work~\cite{hoda2012developing,carver2007use,diaz2023applying,stol2016grounded,adolph2011using,zimmermann2023grounded,foundjem2023grounded} in the domain of empirical software engineering have widely used the GT approach to investigate complex subjects, in terms of developer activities and software engineering practices/strategies. Therefore, we use a GT approach following the previous studies to comprehensively investigate how Chart maintainers mitigate fixable vulnerabilities and why do fixable vulnerabilities remain unfixed. We conduct our GT process based on a set of 90 GitHub repositories that are associated with 686 Helm Charts. We detail the GT repository sampling process in Section~\ref{sec:methodology:subsec:gt_steps_data}, and the iteratively GT coding and analysis process in Section~\ref{sec:methodology:subsec:gt_steps}.

\begin{figure}[t]
    \centering
    \includegraphics[width=\textwidth]{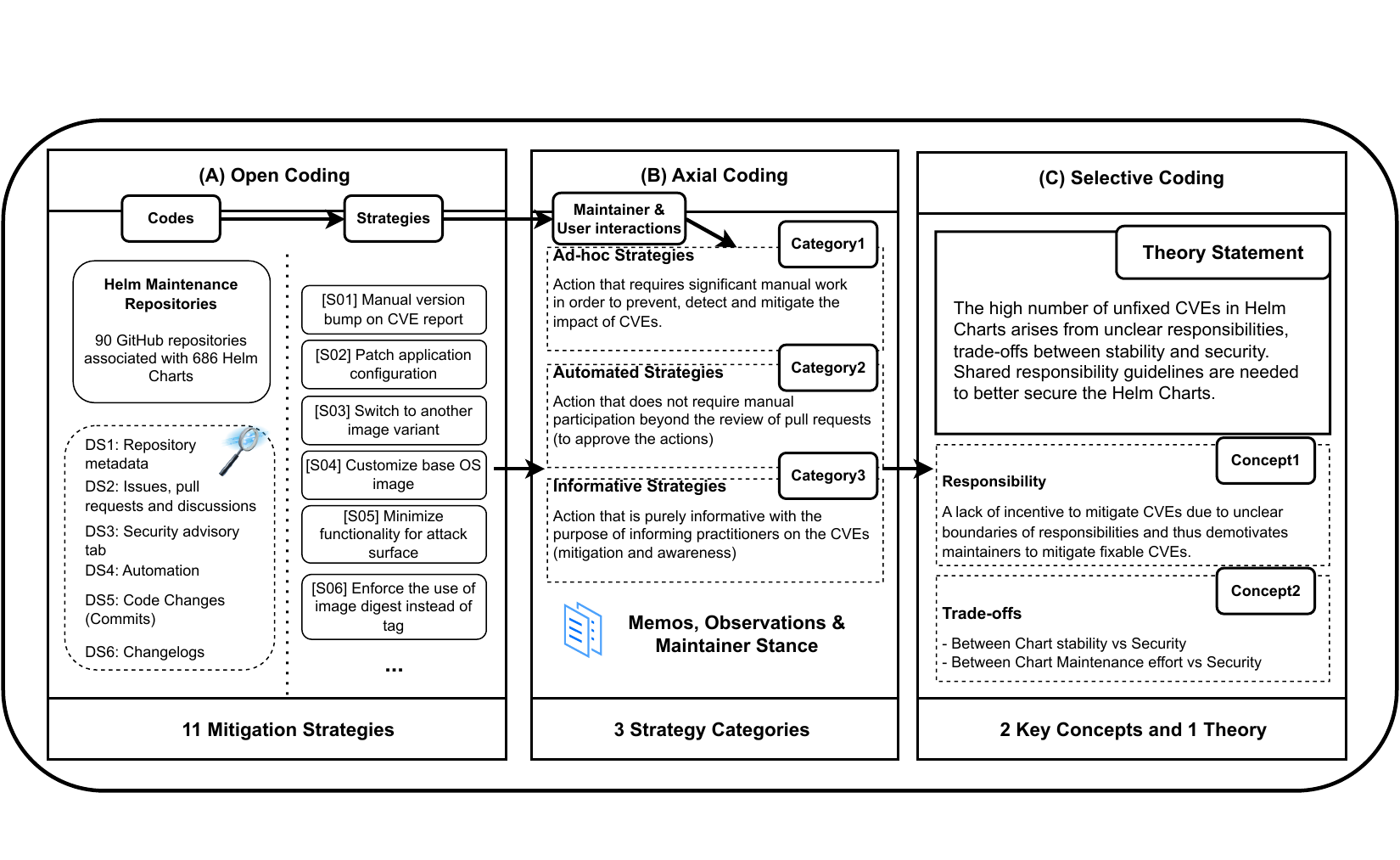}
    \caption{An overview of the Grounded Theory approach, including three coding phases, i.e., Open coding (A), Axial Coding (B), and Selective Coding (C) phases. These phases are conducted iteratively for ten iterations until the first and second authors reach a strong agreement, subsequently confirmed by the third author.}
    \label{fig:grounded}
\end{figure}

We decide to study the maintenance activities of the studied Charts (i.e., 686 Charts) from their associated GitHub repositories (i.e., 90 GitHub repositories) as they contain rich information (as discussed in Section
\ref{sec:background}). We first examine commits history, release notes, issues, pull requests and related discussions around the fixable CVEs. Then, we categorize prevailing approaches and key takeaways of the mitigation strategies. As employing a particular mitigation strategy from a socio-technical perspective is crucial in software maintenance~\cite{mens2016ecosystemic}, we record attempts in maintainers to apply a mitigation strategy but end up with a negative stance (e.g., rejection and unwillingness). We explain the observed application of the strategy and maintainer stance for each identified strategy with a real-world example.

\subsection{Data Sources}
\label{sec:methodology:subsec:data-sources}
We obtain our dataset from two data sources: 1) Artifact Hub, where Helm Charts are hosted, and 2) GitHub repositories, where the associated maintenance activities are recorded. 
During the initial data exploration on Artifact Hub, we identified a large number of publicly available Helm Charts (11,035). Within the dataset, we identified 6,202 Helm Charts with a security report containing an auto-generated list of vulnerabilities affecting the Chart components. We parse the Chart metadata and CVE information within the security reports to perform the quantitative analysis.

As discussed in Section \ref{sec:background}, GitHub repositories contain rich maintenance activities where we could investigate current CVE mitigation strategies and their adoption. In addition, solely relying on the metadata of Chart artifacts and security reports remains insufficient to draw practical conclusions on the prevalence of fixable CVEs that remain unfixed. 

Information related to CVE mitigation strategies is naturally encoded in the GitHub repositories that serve as the source of the Helm Chart releases. Previous studies~\cite{buhlmann2022developers,acar2017security,horawalavithana2019mentions} use GitHub data to understand how developers handle security-related issue reports and provide valuable empirical insights. In our study, we observe that the information related to CVE mitigation exists beyond issue reports but also in CVE-fix-related pull requests, commit messages, automation scripts, and changelogs in GitHub maintenance repositories (see Figure~\ref{fig:grounded}). Such data is naturally suitable for grounded theory exploration, since no existing guidelines could suggest where to mine such strategies~\cite{carver2007use}.

\subsection{Data Collection}
\label{sec:methodology:subsec:data-collection} Figure~\ref{fig:data_collection_procedure} depicts the steps of how we collect the security reports of Helm Charts and their associated GitHub repositories (along with the specific output artifacts from each step). Table~\ref{table:Chart-report-metrics} summarizes the dataset obtained in this study. All scripts and data artifacts involved in the collection, filtering and analysis steps are available in the replication package~\cite{chen_lin_adams_hassan_2023}:

\begin{figure*}[t]
    \centering
    \includegraphics[width=\textwidth]{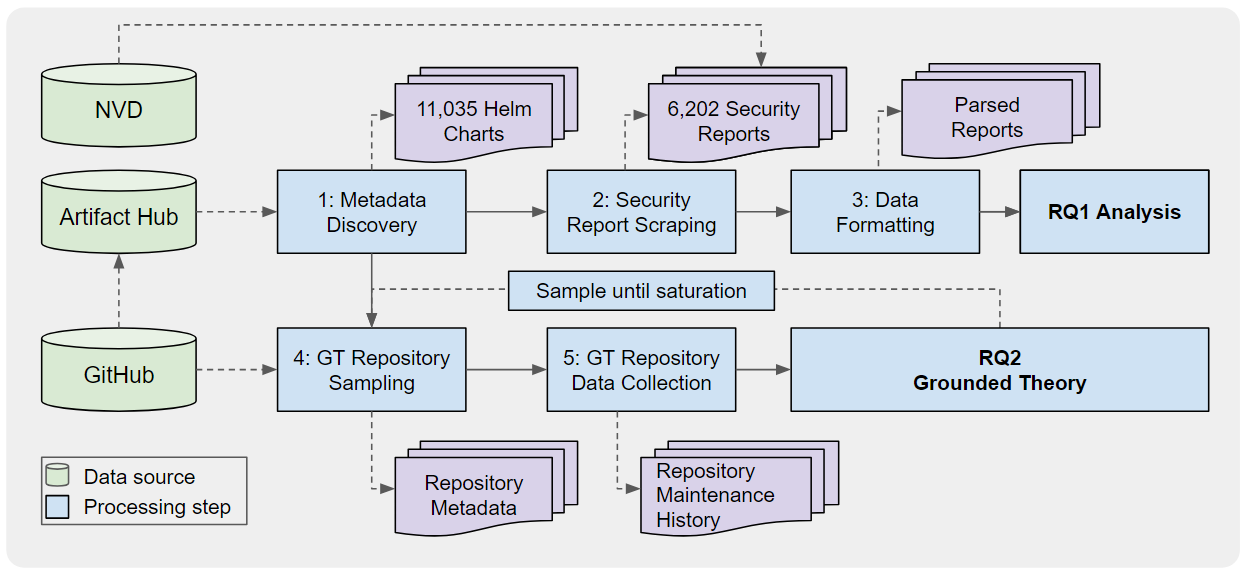}
    \caption{An overview of our data collection steps.}
    \label{fig:data_collection_procedure}
\end{figure*}

\textbf{Step 1: Metadata Discovery:}
We retrieve the metadata (e.g., officiality) of the 11,035 Helm Charts by the Artifact Hub API\footnote{\url{https://artifacthub.io/docs/api/}}, which covers all public Helm Charts published to the Artifact Hub registry since its establishment in 2020\footnote{\url{https://www.cncf.io/projects/artifact-hub/}} until May 2023. We obtained 230 official Charts and 10,805 unofficial Charts.

\textbf{Step 2: Security Report Scraping:} 
According to the package IDs in the metadata of each Helm Chart, we extract its latest security report by the Artifact Hub API\footnote{\url{https://artifacthub.io/docs/api/}}, if available. As a result, we compose a dataset of 6,202 Charts with security reports, i.e., a list of unfixed 13,095 CVEs (see Table \ref{table:unfixed_define}). We further extract the metadata of the 13,095 CVEs from the National Vulnerability Database~\cite{booth2013national} for the analysis in RQ1. 

\textbf{Step 3: Data Formatting:}
For each Chart, we extract metrics required for the two RQs, such as the image and package dependencies of a Helm Chart and the number of affected Charts by each CVE.

\textbf{Step 4: GT Repository Sampling:}
\label{sec:methodology:subsec:gt_steps_data}
As introduced in Section~\ref{sec:background}, Artifact Hub is the registry for Helm Charts, yet the daily development and releasing pipelines of the Charts remain dominantly on GitHub\footnote{\url{https://github.com/}}.
We analyze the associated GitHub repositories of Helm Charts in this study through a GT qualitative study. We detail the GT study data collection in the following paragraphs.

During the initial data exploration phase, we observed that the source code of many Helm Charts is hosted on GitHub repositories that are named differently from the Chart. Therefore, it is difficult to backtrack all the Charts to their associated GitHub repositories accurately. We extract the associated repositories by the Artifact Hub metadata API\footnote{\url{https://artifacthub.io/docs/api/}}, if applicable. We obtained a set of 3,486 Helm Charts that could be backtracked to their associated GitHub repositories.

We follow the theoretical sampling methodology and guidelines~\cite{glaser2017theoretical,baltes2022sampling} to sample GitHub repositories and associated Helm Charts for the GT investigation. Previous empirical software engineering studies have applied theoretical sampling to systematically derive insights from complex data (i.e., software maintenance, evolution, practices) until theoretical saturation~\cite{baltes2022sampling,wohlin2015towards,adolph2011using,seaman1999qualitative}. If the saturation is not reached after the three coding phases, this dataset will be extended. In other words, saturation occurs when further data collection and analysis reaches diminishing results in terms of new strategies and strategy observations~\cite{hoda2012developing,zimmermann2023grounded}. In our study, the theoretical saturation indicates that the emerged concepts and observations for CVE mitigation strategies could be formed into a GT with sufficient evidence, and we do not identify any new concepts in a new sample from the GitHub repositories, similar to 
 prior work~\cite{rowlands2016we,adolph2011using,pano2018factors}. 

At the end of our GT coding phases (which will be introduced in Section~\ref{sec:methodology:subsec:gt_steps}), we have utilized information from 90 GitHub repositories. The first and second authors extended repository samples based on the coding results from the previous sample (e.g., look for repositories with diverse security policy descriptions to understand CVE information disclosure). We decided our initial sample to be Bitnami Charts\footnote{\url{https://github.com/bitnami/charts}}, which is a repository that contains 105 Charts and does not have any fixable CVEs in all releases. We consider the Bitnami Charts repository as a reasonable starting point to bootstrap the list of strategies, which evidently ensure the mitigation of all fixable CVEs in Charts before releases.

\textbf{Step 5: GT Repository Data Collection:} 
Upon gathering the 3,486 Charts associated with GitHub repositories, we conduct a quality check based on the repository metadata before taking a sample. We ensure the quality of each sampled repository by following common guidelines of Mining Software Repositories (MSR)~\cite{hassan2008road,dabic2021sampling}. Specifically, we apply a quantitative filter for repositories to ensure each analyzed repository is not a fork (except template repositories) or mirror. We ensure that the repositories have released more than one Chart version, providing evidence for meaningful development activities. Since we expect to explore as many mitigation strategies as possible via the GT approach, we analyze various data types in a repository. In other words, we do not only study specific types of data (e.g., a minimal amount of issues and pull requests) since other data types (e.g., automation and release notes in the repository) could also provide valuable insights.

In the open coding phase, We identify vulnerability-related issue reports, commit messages, and pull requests by searching for a collection of keywords connected by the Boolean OR condition, including ``CVE'', ``GHSA''\footnote{\url{https://github.com/advisories}}, ``NSWG''\footnote{\url{https://github.com/nodejs/security-wg}}, ``SUSE''\footnote{\url{https://www.suse.com/security/cve/}}, and ``vulnerability(ies)''. The keywords (i.e., GHSA, NSWG and SUSE) are included since we observe that several CVEs were initially assigned with an alternative alias according to their origin (GitHub, NodeJS and SUSE Linux distribution).

\subsection{Analysis Steps} 
\label{sec:methodology:subsec:steps}
\begin{figure*}[t]
    \centering
    \includegraphics[width=\textwidth]{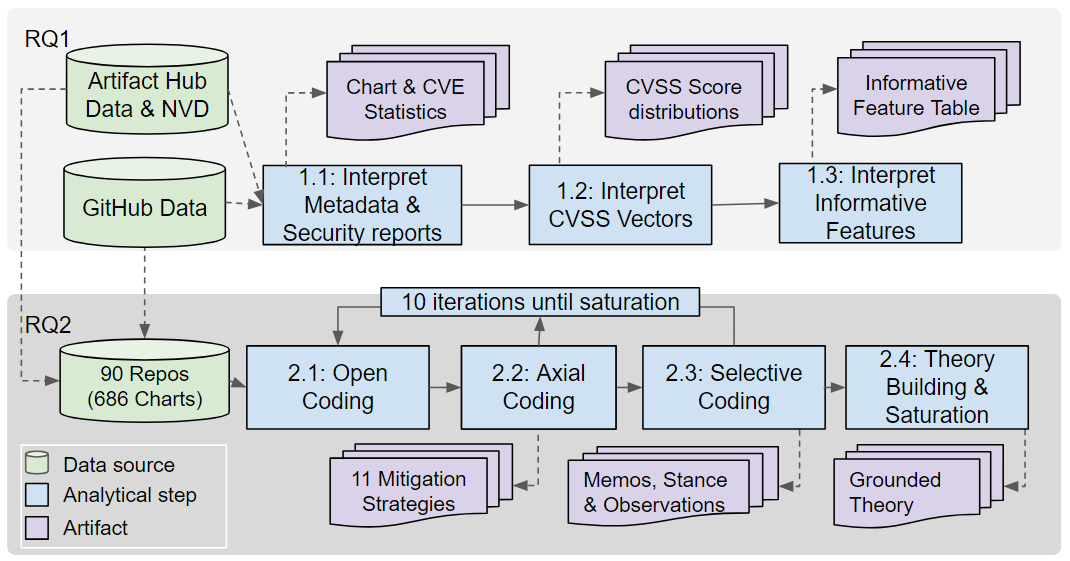}
    \caption{An overview of our study design.}
    \label{fig:study_protocol}
\end{figure*}

This section introduces the study protocol for the analyses in each RQ. Figure~\ref{fig:study_protocol} depicts the specific output artifacts (purple) in each analysis step of the 2 RQs (blue) and the data source (green) where the data is obtained, as introduced in the previous section.

\textbf{Step 1.1: Interpret Metadata and Security Reports:}
Before diving into the details of fixable CVEs that remain unfixed in Helm Charts, we analyze the Helm Charts landscape with high-level metrics in order to understand the prevalence of fixable vulnerabilities. We use the rich metadata shipped in the latest release of the 11,035 Helm Charts, including their dependencies and CVE information. We first calculate the number of unique CVEs that are fixable in proportion to the total number of unfixed CVEs. Then, we rank a top 20 common set of fixable CVEs by the number of affected Charts. Step 1.1 produces the ``Chart \& CVE Statistics'' artifact, as shown in the figure.

\textbf{Step 1.2: Interpret CVSS Vectors}
We measure the severity of the CVEs with the CVSS V3 base score and CVSS V3 vector, which are the standard for measuring the general severity of the CVEs~\cite{first2015common} in the Helm Charts ecosystem. We use the CVSS V3 metrics (i.e., 8 metrics, as introduced in Section~\ref{sec:sub_cvss}) of each CVE to calculate corresponding CVSS V3 base scores, exploitability scores and impact scores. 

Subsequently, we aggregate the Helm Charts based on the officiality metadata on Artifact Hub, which is an indicator of whether the core dependencies of a Helm Chart are maintained by the same organization developing those. We conduct non-parametric statistical tests to measure whether the plotted distribution split by officiality shows statistically significant differences. We use the Mann-Whitney U Test, followed by calculating Cliff's Delta to measure the practical effect size of the observation~\cite{nachar2008mann}. Since CVSS V3 scores are numerical and are assigned with a value ranging from 0 to 10, we use a split violin plot to visualize the distributions when aggregated by officiality. Step 1.2 produces the ``CVSS Score Distributions'' artifact.

\textbf{Step 1.3: Interpret Informative Features:}
We investigate the utilization of the security features provided by Artifact Hub and GitHub, which include:

\begin{itemize}
    \item Automatic security scanner reports generated by Trivy (the source of 6,202 security reports on Artifact Hub)
    \item An indicator for security-related updates in a release (Artifact Hub)
    \item Security advisory (GitHub)
    \item Security policy, defined by ``SECURITY.md"(GitHub)
    \item Release changelogs (GitHub)
\end{itemize}

We determine the utilization data by counting the number of Helm Charts that enable such a feature and populate corresponding data (CVE information). The purpose is to understand whether such features are widely used to inform Chart users of the fixable vulnerabilities in Helm Charts that remain unfixed. Step 1.2 produces the ``Informative Feature Table'' artifact.

\label{sec:methodology:subsec:gt_steps}
\textbf{Step 2.1: Open Coding (Iterative):}
As introduced in Section~\ref{sec:methodology:subsec:gt_steps_data}, we conduct the 
GT process through theoretical sampling. Due to the nuanced nature of data in the Helm Charts domain, we conduct 10 iterations of each coding phase that will be introduced in Steps 2.1, 2.2 and 2.3. 

Figure~\ref{fig:grounded} presents a detailed view of our GT study and corresponding results. We follow the methodology used by previous GT studies~\cite{foundjem2023grounded,diaz2023applying} to ensure strong inter-rater agreement after conducting 10 coding iterations. Both the first and second authors met frequently in person to discuss the coding results, memos, observations and concepts (after each coding iteration). We end the coding iterations when a strong agreement has been reached after sampling and analyzing the 90 GitHub repositories associated with 686 Helm Charts~\cite{diaz2023applying}. Upon the coding results being discussed and formulated into writing, they are subsequently confirmed by the third author to provide unbiased opinions. 

In the first iteration of open coding, we investigate all data types, including issues, pull requests, source code, continuous integration runs and release notes to extract any strategies applied to mitigate CVEs from our initial repository (i.e., Bitnami Charts). Specifically, we manually investigate the following data:

\begin{itemize}
    \item Repository metadata (READMEs, Links to Documentation): Gain insights on the potential mitigation strategies for unfixed vulnerabilities applied in the repository.
    \item Issues, pull requests and discussions: Specific discussions and changes attempting to mitigate CVEs, maintainer stance over such changes and proposals.
    \item Security advisory tab: Repository policy on CVE disclosure and mitigation.
    \item Automation (i.e., GitHub Actions\footnote{\url{https://github.com/features/actions}}): Automation definition and results to detect CVEs and apply automatic patches.
    \item Code Changes (Commits): Detailed Chart manifest changes to release.
    \item Changelogs: Changes between releases, mention of CVE fixes and notes.
\end{itemize}

For a given repository, we write down memos of existing mitigation strategies and count the usage of such a strategy. We add a strategy to the list if it is unseen before. After each coding iteration, the first and second authors discuss the collected data and the individually writes down memos before a discussion. Through later iterations, we gradually merge strategies if they present similar actions, and we write down details regarding each strategy to prepare the basis of empirical insights.

Throughout this process, we also note whether the mitigation strategy applies to all situations or whether the maintainers expressed mixed opinions, including rejection or hesitation. For example, some maintainers are against using CVE scanners due to false positives. We record other observations that could cross findings with previous RQs, such as the lack of use of informative security features.

\textbf{Step 2.2: Axial Coding (Iterative):}
We follow the guidelines of GT~\cite{stol2016grounded} to group the strategies identified in Step 2.1 into categories. We refine the categories iteratively through our memos to develop finite categories and assign the strategies to each based on the strategy characteristics. Based on the open coding results where we record the occurrence of strategies in each repository, we revisit the strategies to confirm that our observations remain up-to-date after the previous full iteration. We consider this step completed when all strategies could be fit into a category and finalized with an accurate description, which took approximately 10 iterations of coding and discussions by the first two authors and finally confirmed by the third author. Step 2.2 produces the ``Mitigation Strategies'' artifact.

\textbf{Step 2.3: Selective Coding:}
We focus on the core categories identified in the axial coding phase (Step 2.2). We draw interconnections between categories, focusing on the mitigation strategies employed by maintainers for CVEs.

We construct a theoretical draft by integrating and refining the categories and their properties around this central phenomenon. We build a narrative around how the discovered mitigation strategies are used, along with their conditions and consequences, with practical examples on GitHub. This step yields the ``Strategy Stance \& Observations'' artifact.

\textbf{Step 2.4: Building the Theory:}
We synthesize insights from the selective coding process (Step 2.3) to develop a grounded theory. The constructed GT explains the maintainers' strategies in terms of how maintainers mitigate CVEs for Helm Charts. At this stage, we propose an actionable theory and recommendations that guide future researchers and practitioners interested in improving Helm Charts' security. 
We also bring extra insights to observations that could not be explained via the quantitative results in RQ1. Step 2.4 results in the ``Grounded Theory'' artifact.

\section{RQ1 results: \RQOne} 
\label{sec:rq1-results}

\subsection{Prevalence of Helm Charts Vulnerabilities}
\label{sec:rq1-results:subsec:cve_prevalence}

\textbf{13,095 unique CVEs affect across the latest versions of the 6,202 Helm Charts (with a security report), and 10,982 (83.9\%) of these CVEs are fixable} (See Table~\ref{table:unfixed_define}). In particular, the 6,202 Helm Charts are affected by a median of 119 unfixed CVEs (i.e., the union set of fixable and not-patch-yet CVEs) and a median of 52 (43.7\%) fixable CVEs. Only 42 (i.e., 0.7\%) of the 6,202 Helm Charts are not vulnerable to any CVEs, and 439 (i.e., 7.1\%) are not vulnerable to any fixable CVEs. 

Though attackers may not exploit a vulnerability, \citet{dullien2017weird} highlighted that it is inherently difficult to prove that the vulnerability would not be exploited in a complex system like Kubernetes, even for experienced engineers. Due to such difficulty, the large number of unfixed CVEs in Helm Charts could lead to severe security consequences, such as frequent false alerts from security audits\footnote{\url{https://github.com/kubernetes/ingress-nginx/issues/8520}} and potential malicious attacks\footnote{\url{https://nvd.nist.gov/vuln/detail/CVE-2021-44228}}.

\textbf{The majority (84.6\%, 9293) of fixable CVEs affect more than one Chart, and the top fixable CVE (ranked by the number of affected Helm Charts) affects 2,517 (40.5\% of 6,202) Charts. }
Table~\ref{table:cve-by-frequency} presents the top 20 fixable CVEs sorted by the number of affected Helm Charts. Among the top 20 CVEs, 2 (10\%) is of critical severity level, 11 (55\%) are of high severity, and 7 (35\%) are of medium severity. The top 20 fixable CVEs affect at least 1,300 ($>$20.9\%) of the studied 6,202 Helm Charts. Such a large number of Helm Charts affected by a small set of 20 CVEs suggests that maintainers across the affected Charts could work together to mitigate the CVEs faster, compared to working independently by the maintainers of one particular Chart. 

\begin{table}[t]
\centering
\caption{The top 20 fixable CVEs ranked by \#affected Helm Charts.}
\label{table:cve-by-frequency}
\begin{tabular}{@{}p{2.8cm}rcp{2.8cm}rc@{}}
\toprule
\textbf{CVE ID} & \textbf{\#} & \textbf{Severity$^*$} & \textbf{CVE ID} & \textbf{\#} & \textbf{Severity$^*$} \\
\midrule
CVE-2023-0286 & 2,517 & H & CVE-2023-0464 & 1,638 & H \\
CVE-2023-0215 & 2,410 & H & CVE-2022-29458 & 1,600 & H \\
CVE-2022-4304 & 2,410 & M & CVE-2022-29526 & 1,574 & M  \\
CVE-2022-4450 & 2,403 & H & CVE-2022-42898 & 1,569 & H \\
CVE-2022-37434 & 2,145 & C & CVE-2023-23916 & 1,457 & M \\
CVE-2018-25032 & 2,007 & H &  CVE-2022-41723 & 1,438 & H \\
CVE-2022-0778 & 1,870 & H & CVE-2022-4415 & 1,405 & M \\
CVE-2022-1271 & 1,678 & H & CVE-2021-3712 & 1,366 & M \\
CVE-2023-0361 & 1,655 & H & CVE-2022-41717 &1,362 & M \\
CVE-2023-0465 & 1,644 & M & CVE-2022-1292 & 1,347 & C \\
\bottomrule
\end{tabular}
\raggedright $^*$ Severity representation: L=Low, M=Medium, H=High, C=Critical \\
\end{table}

Interestingly, the majority (14, 70\%) of the top 20 fixable CVEs were disclosed before 2023, with the oldest (CVE-2018-25032) dating back to 2018. This suggests that these fixable CVEs remain unfixed for a prolonged period, i.e., at least one year. As new functionalities are frequently introduced in a given Helm Chart by leveraging other Charts (i.e., Chart dependencies) and/or images and packages (i.e., image/package dependencies), the Helm Chart could become vulnerable due to vulnerable Chart dependencies and/or images, packages being used.

The existence of fixable CVEs (which may or may not have actual impact) in Helm Charts could introduce additional informational overhead~\cite{rombaut2023there} to the Chart users. Such CVEs could lead to misunderstanding over the integrity of the Helm Charts. As CVE scanners are prone to false positives~\cite{mirhosseini2017can}, maintainers struggle with dealing with false alarms, which may lead to disabling such CVE scanners. On the other hand, Chart users often rely on such scanning results due to the tight security requirements of large organizations~\footnote{\url{https://github.com/amcharts/amcharts5/issues/897#issuecomment-1536335897}}.

\begin{figure}[t]
    \centering
    \begin{subfigure}{0.32\textwidth}
        \centering
        \includegraphics[width=\linewidth]{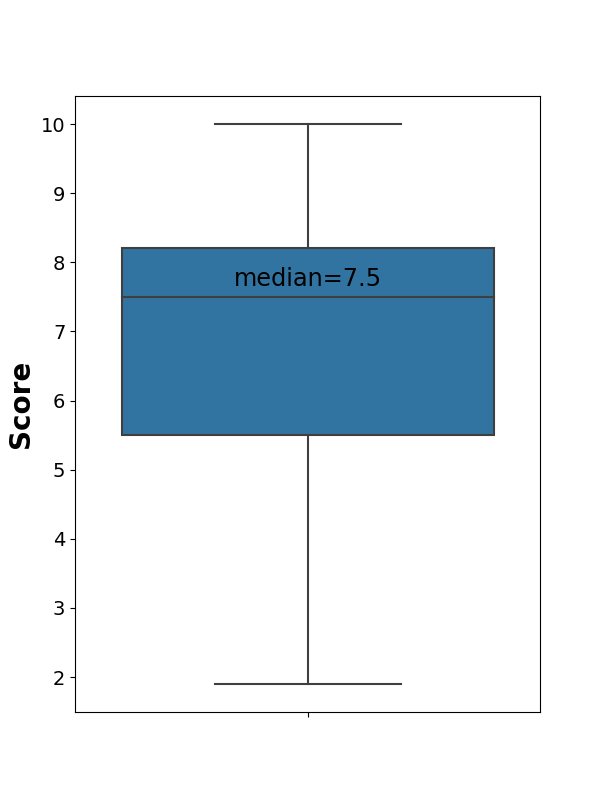}
        \subcaption{Base Score.}
        \label{fig:overall_cvss}
    \end{subfigure}
    \hfill 
    \begin{subfigure}{0.32\textwidth}
        \centering
        \includegraphics[width=\linewidth]{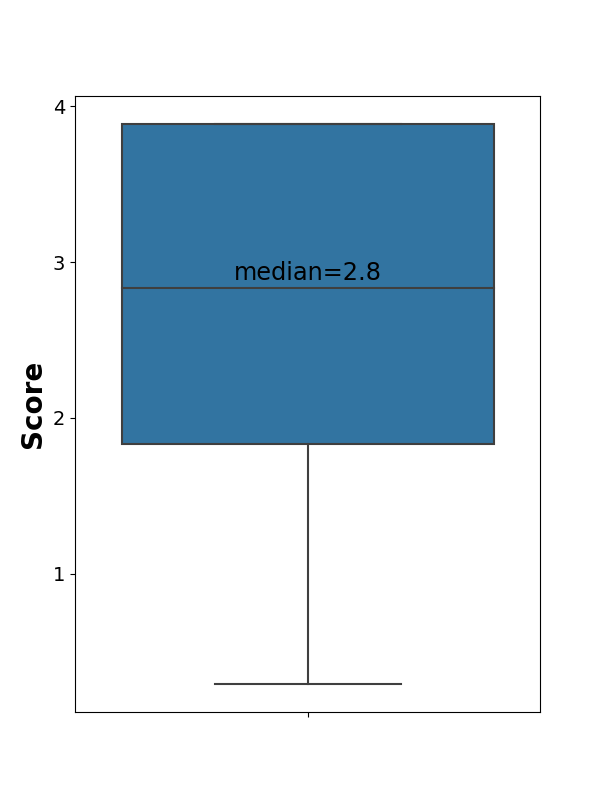}
        \subcaption{Exploitability score.}
        \label{fig:overall_exploit}
    \end{subfigure}
    \hfill 
    \begin{subfigure}{0.32\textwidth}
        \centering
        \includegraphics[width=\linewidth]{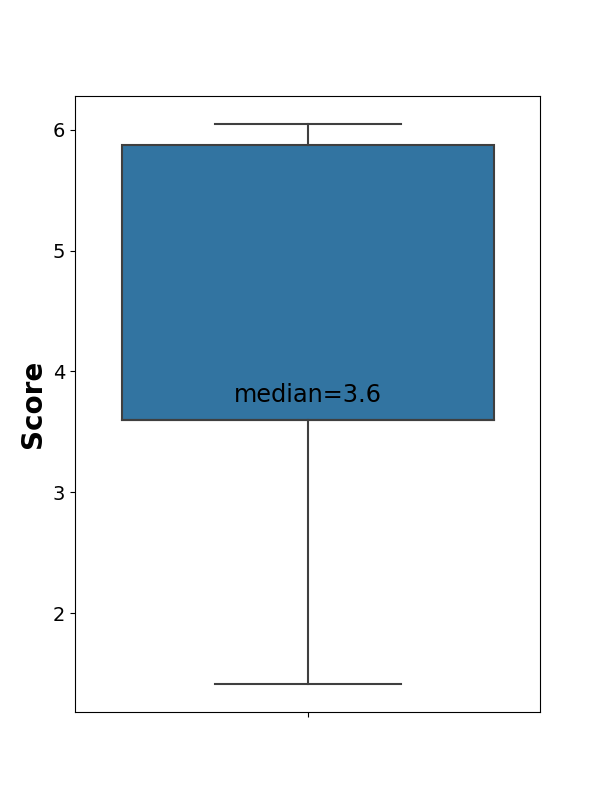}
        \subcaption{Impact score.}
        \label{fig:overall_impact}
    \end{subfigure}
    \caption{CVSS V3 scores of 10,982 fixable CVEs in the 6,202 Helm Charts, broken down by CVSS base score (a), exploitability score (b), and impact score (c). The majority of the fixable CVEs are at \textbf{high} severity level (i.e. base score of 7.0-8.9). The fixable CVEs are \textbf{easily exploitable} (i.e., median of 2.8 out of 3.9), though with moderate impact (i.e., median of 3.8 out of 6.0). }
    \label{fig:metrics-overall}
\end{figure}

\begin{figure}[t]
    \centering   
    \begin{subfigure}{\textwidth}
        \centering
        \includegraphics[width=\textwidth]{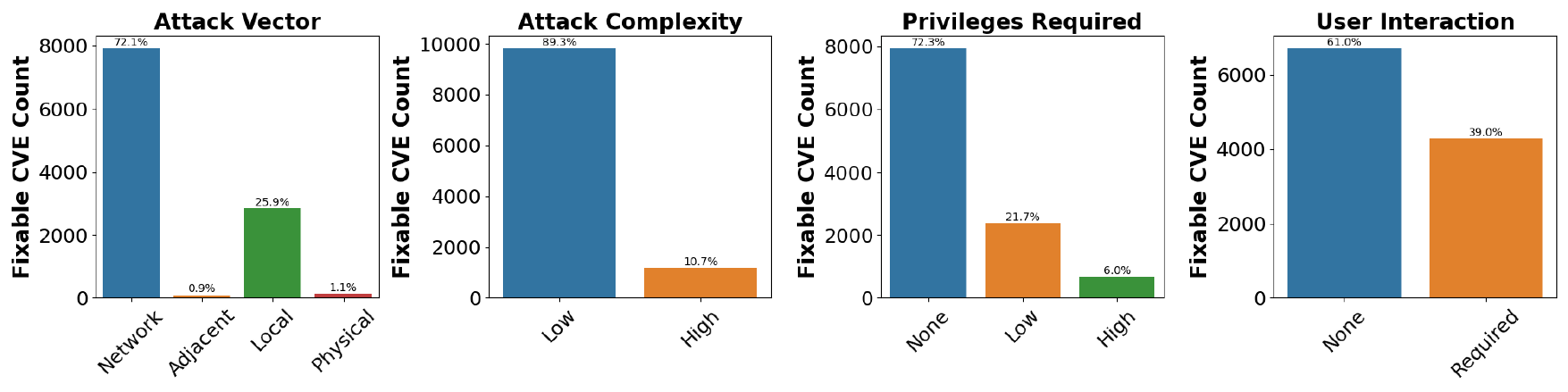}
        \caption{Exploitability metrics breakdown.}
        \label{fig:exploitability-breakdown}
    \end{subfigure}
    \vspace{5mm}
    \begin{subfigure}{\textwidth}
        \centering
        \includegraphics[width=\textwidth]{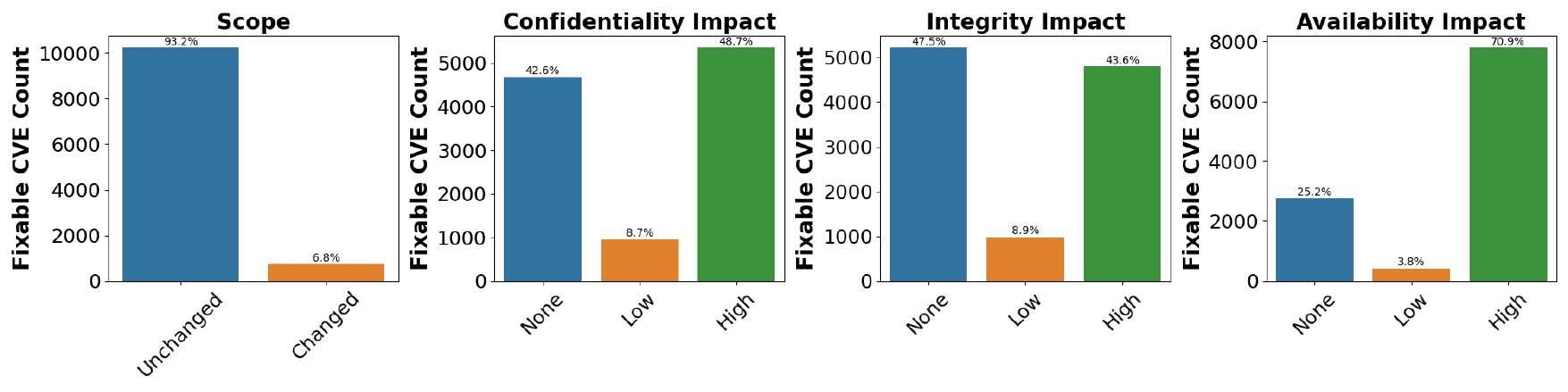}
        \caption{Scope and impact metrics breakdown.}
        \label{fig:impact-breakdown}
    \end{subfigure}
    \caption{CVSS V3 metrics of the fixable CVEs in the 6,202 Helm Charts, broken down by exploitability (a) scope and impact metrics (b).}
    \label{fig:metrics-breakdown}
\end{figure}

\begin{figure}[!htbp]
    \centering
    \includegraphics[width=\textwidth]{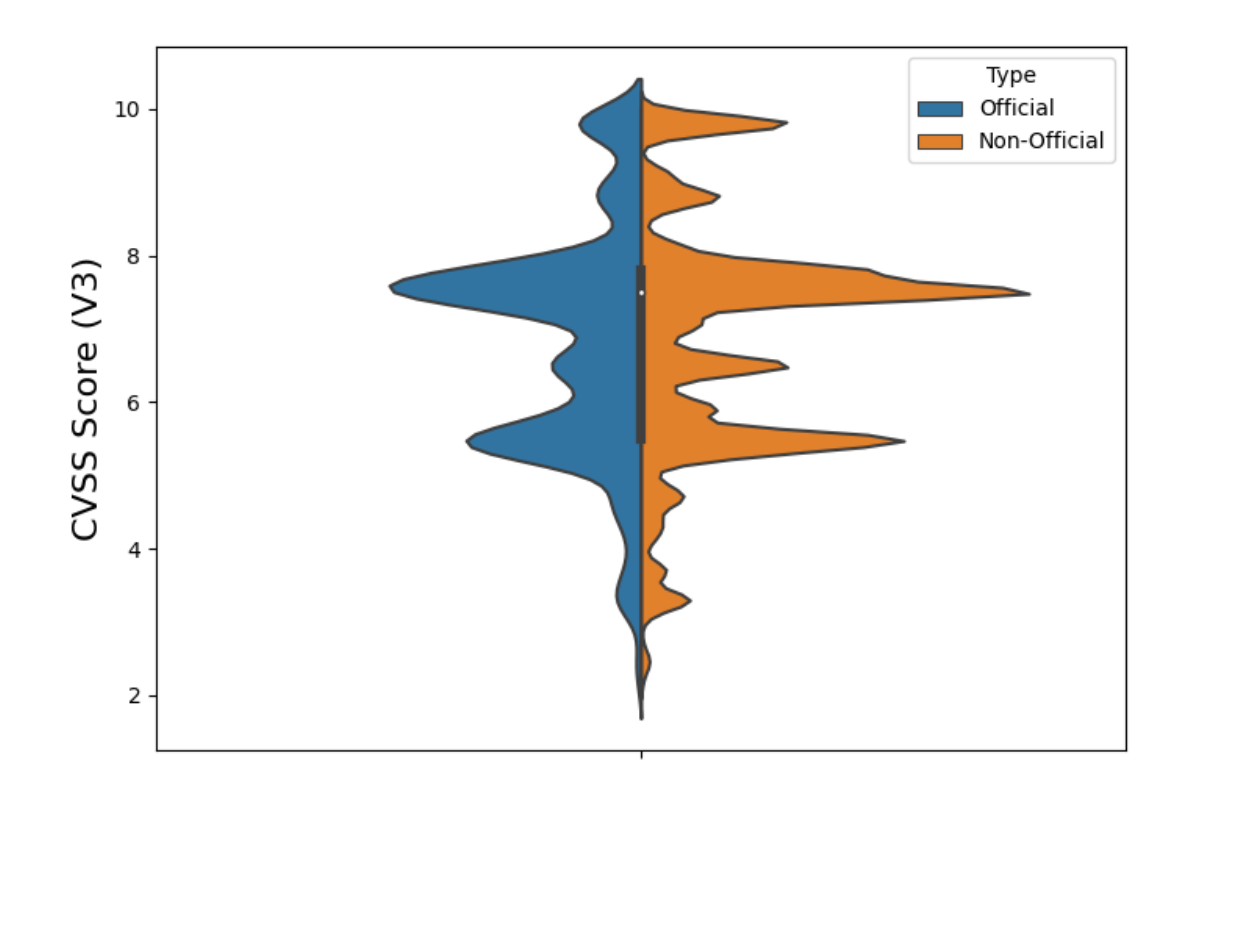}
    \caption{The distribution of CVSS V3 Score of Fixable CVEs that affect official vs. unofficial Charts. Fixable CVEs that affect both official and unofficial Charts have a median of \textbf{high} severity, i.e., a median of a base score of 7.3 and 7.5, respectively.}
    \label{fig:cve_distribution_violin_official}
\end{figure}

\subsection{CVSS V3 Scores and Metrics}

\textbf{The 10,982 fixable CVEs have a median severity of ``high" (i.e., a median score of 7.5 out of 10). According to the CVSS V3 specification, the CVEs can be easily exploited (i.e., a median score of 2.8 out of 3.9), though with a moderate impact (i.e., a median score of 3.6 out of 6.0).}

Figure~\ref{fig:metrics-overall} depicts the distribution of CVSS V3 base, exploitability, and impact scores (ref Section \ref{sec:background}). We detail the 8 metrics in the exploitability and impact groups in Figures~\ref{fig:exploitability-breakdown} and~\ref{fig:impact-breakdown}. The fixable CVEs in the Helm Charts are risky as they are more likely to be exploited, i.e., the attack complexity of 89.3\% (9,809) of the CVEs is low, exploiting 72.3\% (7,939) does not require any privileges and attackers do not need to gain access to a particular network, as shown in Figure~\ref{fig:exploitability-breakdown}.

\textbf{There exists a significant difference between fixable CVEs that affect official vs. unofficial (p-value $<$ 0.001), with a negligible effect size of -0.0003 (Cliff's $\delta<0.147$).} Figure~\ref{fig:cve_distribution_violin_official} visualizes the comparison of the distributions of the CVSS score of fixable CVEs that affect official vs. unofficial Helm Charts. Mann-Whitney U test reports significant differences between official and unofficial Charts with p-value $<$ 0.001. However, the negligible effect size of -0.0003 (Cliff's $\delta<0.147$~\cite{gupta2017impact}) indicates that practitioners may not prefer to use official Charts over unofficial Charts in terms of Chart security. 

Our result is different from prior works~\cite{wist2021vulnerability, ibrahim2021study} that studied vulnerabilities in the Docker ecosystem. \citet{wist2021vulnerability} reported that official images on the Docker Hub are more secure than unofficial ones regarding the high-to-critical severity (i.e., a CVSS score ranging from 7.0 to 10.0) vulnerabilities. \citet{ibrahim2021study} reported that official images would attract more usage in Docker Compose files. One possible reason for such differences in using official images and Charts is that the two ecosystems are of different maturity. While Docker images and Helm Charts are both related to containerization, the Docker Hub ecosystem\footnote{\url{https://hub.docker.com/}}, being a well-established registry for Docker images, exhibits a different security landscape, compared to Helm Charts on Artifact Hub, which is still in the incubation phase supervised by the CNCF (as discussed in Section \ref{sec:background}). In addition, Docker Hub is maintained by a commercial company, while the CNCF is a non-profit organization driven by the community.

\subsection{Informative security features}
\label{sec:rq1-results:subsec:informative_features}
\textbf{44\% (4,833) of the studied 11,035 Helm Charts disable security reports.} Having a security audit is a project graduation criterion employed by the CNCF (as discussed in Section \ref{sec:background}). The requirement for having a security audit provides project maintainers with an immediate overview of the risks for adversarial attacks, and encourages continuous detection of vulnerabilities\footnote{\url{https://www.cncf.io/blog/2022/08/08/improving-cncf-security-posture-with-independent-security-audits/}}. In fact, there exist many tools, such as Trivy\footnote{https://github.com/aquasecurity/trivy}, that assist Chart maintainers effortlessly generate a security report periodically.

\textbf{Although 56\% (6,202) of Helm Charts disclose a security report, most do not notify their users of sufficient information related to mitigating vulnerabilities.} In particular, only 80 Charts (1\%) explicitly indicate the vulnerability fixes in their changelogs by a ``\textit{has\_security\_fix}'' label. The label indicates whether a Chart release incorporates a CVE fix, which is one of the best security practices on Artifact Hub\footnote{\url{https://artifacthub.io/docs/topics/annotations/helm/}}.

\begin{table}[t]
\centering
\caption{The utilization of informative features with respect to officiality. Metrics include \#Charts that disclose a security advisory, \#Charts that follow a security policy, \#Charts that indicate a CVE fix at least once in their changelogs, and median/mean of releases incorporating CVE fixes per Chart.}
\begin{tabular}{@{}p{5cm}rr@{}}
\toprule
Metric & Official (170 Charts) & Unofficial (3316 Charts) \\ \midrule
Security advisory  & 20 (11.8\%) & 421 (12.7\%)\\
Security policy  & 41 (24.3\%) & 443 (13.3\%) \\ 
Changelog (Fixes CVEs at least once)  & 55 (32.5\%)  &  367 (11.1\%) \\
Changelog (Median/Mean) & 1/4.7 & 0/3.1  \\
\bottomrule
\end{tabular}
\label{table:informative-metrics-github}
\end{table}

Table~\ref{table:informative-metrics-github} presents the utilization of informative features about the 3,486 Helm Charts (i.e., Charts of which we found the associated GitHub repositories). We observe that Chart maintainers typically do not utilize such informative features in the Chart-associated GitHub repositories. Specifically, maintainers of 20 (11.8\%) official and 421 (12.7\%) unofficial Charts utilize a security advisory. Security advisory is a GitHub built-in feature\footnote{\url{https://docs.github.com/en/code-security/security-advisories/}} that maintainers could disclose the information related to fixed CVEs. As a result, users could use such information as a guideline to upgrade Helm Charts and prevent security threats. Maintainers of 41 (24.3\%) official and 443 (13.3\%) unofficial Charts have a security policy. Security policy presents the contents of a special \textit{SECURITY.md} file hosted in the Helm Chart repositories. This file is intended to provide guidelines for reporting security vulnerabilities\footnote{\url{https://docs.github.com/en/code-security}}. In terms of changelog documentation, 55 (32.5\%) official Charts and 367 (11.1\%) unofficial Charts mention a CVE fix at least once. The median and mean numbers of releases mentioning CVE fixes are 1 and 4.7 for official Charts, and 0 and 3.1 for unofficial Charts, respectively.

\subsection{Image, Package Dependencies in Helm Charts}
\label{sec:rq1-results:subsec:cve_distribution}

\textbf{The 56\% (6,202) of Helm Charts are composed of a median of 1 Chart dependency, a median of 156 package dependencies, and a median of 1 image dependency.}
We observe that the complexity of Helm Charts varies significantly in the dataset in terms of image counts (i.e. a variance of 31). Due to the nested structure of Helm Charts as discussed in Section \ref{subsec:background-charts-dependencies}, while a typical (median) Helm Chart may only contain one image, it often contains more than 156 package dependencies that are built into the specified image. 

Similar to Docker images, Chart maintainers frequently use the parent-child interconnection in Helm Charts (i.e., Chart dependencies) when deploying applications. \citet{zerouali2022helm} also described a similar phenomenon as the authors find more than 50\% of the Helm Charts only deploy a container image while running multiple containers that are based on the specified image and the specified configurations of the Helm Charts. 

\begin{figure}[t]
    \centering   
    \begin{subfigure}[b]{.32\textwidth}
        \centering
        \includegraphics[width=\linewidth]{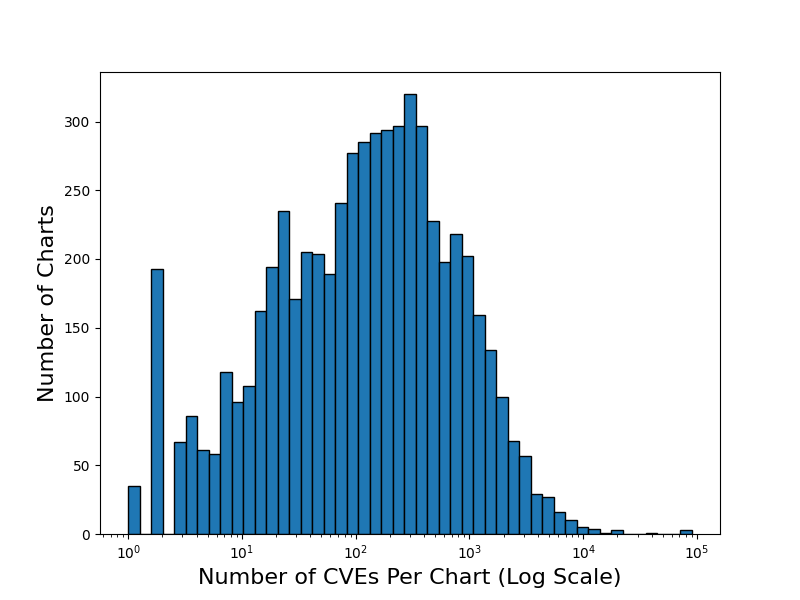}
        \subcaption{Fixable CVEs per Chart.}
        \label{fig:cve_distribution}
    \end{subfigure}
    \begin{subfigure}[b]{.32\textwidth}
        \centering
        \includegraphics[width=\linewidth]{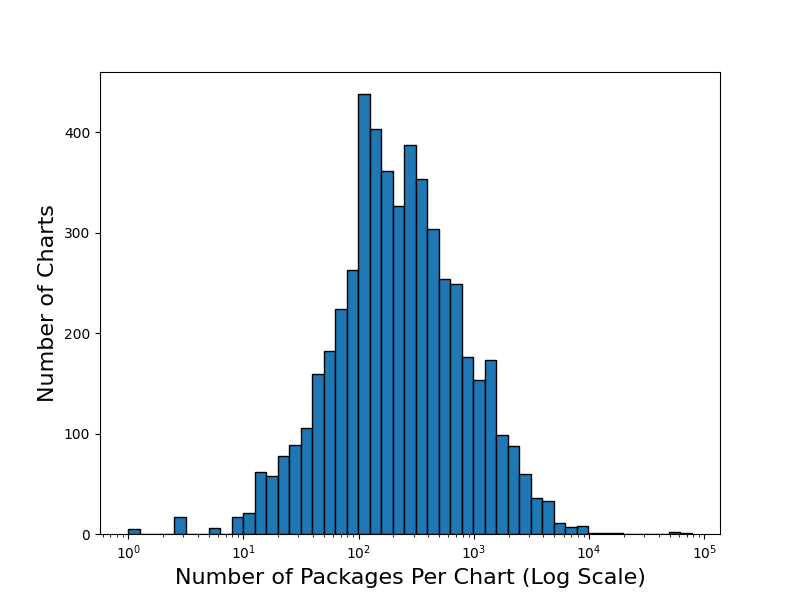}
        \subcaption{Packages per Chart.}
        \label{fig:pkg_distribution}
    \end{subfigure}
    \begin{subfigure}[b]{.32\textwidth}
        \centering
        \includegraphics[width=\linewidth]{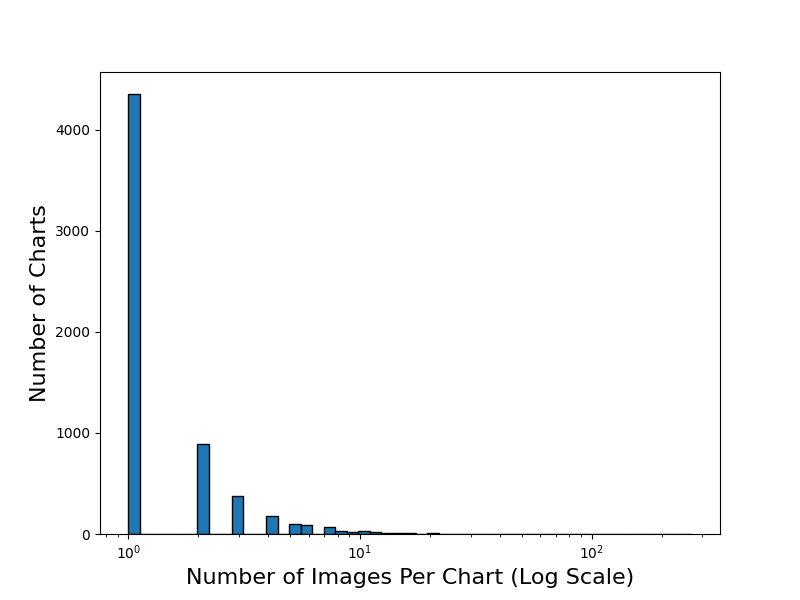}
        \subcaption{Images per Chart.}
        \label{fig:image_distribution}
    \end{subfigure}
    \caption{Distribution of fixable CVE counts (a), dependency package counts (b), and image counts (c) in the 6,202 Helm Charts.}
    \label{fig:combined_distribution}
\end{figure}

\textbf{The number of package dependencies significantly correlates with the number of fixable CVEs (p\(<\)0.001), while the number of images does not.} Figures~\ref{fig:cve_distribution},~\ref{fig:pkg_distribution}, and~\ref{fig:image_distribution} depict the distributions of the number of CVEs, package dependencies, and images involved in the Helm Charts, respectively, at the log scale. We observe a high variance in the number of fixable CVEs and package dependencies in Helm Charts, ranging from one to extreme values. There are 13 Helm Charts, each affected by more than 10,000 vulnerabilities (with some duplication, a CVE may affect multiple package/image dependencies in the Chart). For example, there are 14,602 vulnerabilities in \textit{ScienceBox}\footnote{https://artifacthub.io/packages/helm/sciencebox/sciencebox}, a Helm Chart that simplifies the deployment of services for nuclear science research and contains 17 images. Specifically, 1348 (i.e., 144 critical and 1244 high severity) fixable CVEs in the \textit{ScienceBox} Chart are of high to critical severity (i.e., CVSS V3 score $>$ 7.0).

\begin{table}[t]
  \centering
  \caption{Skewness and normality test results For fixable CVEs, package and image dependencies in 6,202 Helm Charts}
  \label{tab:skewness-normality}
  \begin{tabular}{@{}llll@{}}
    \toprule
    \textbf{Metric} & \textbf{Skewness} & \textbf{Shapiro-Wilk Test} & \textbf{Anderson-Darling Test} \\
    \midrule
    cve\_per\_Chart & 26.6 & 0.17 & 1326.0 \\
    package\_per\_Chart & 29.6 & 0.17 & 1098.7 \\
    image\_per\_Chart & 37.0 & 0.10 & 1605.3 \\
    \bottomrule
  \end{tabular}
\end{table}

Table~\ref{tab:skewness-normality} presents the skewness coefficients and results of the normality tests (Shapiro-Wilk Test and Anderson Darling Test~\cite{razali2011power}) for the three distributions (i.e., fixable CVEs per Chart, package per Chart and image per Chart). Overall, all three metrics (i.e., 26.6, 29.6, 37.0) indicate high skewness, providing evidence to reject the null hypothesis of normality. In particular, we observe that the number of images per Chart demonstrates a significantly higher skewness coefficient of 37.0 than that of package count (29.6). Based on this, we conduct a partial correlation test with the number of package dependencies and images as independent variables and the number of fixable CVEs in a Chart as the dependent variable.

The partial correlation coefficients of 0.85 and -0.47 (p\(<\)0.001) support our observation that the high number of fixable CVEs involved in a Chart does not necessarily correlate to the number of images deployed by a Chart. Instead, it significantly correlates to the number of package dependencies in a Chart. As a container image could consist of a large number of layers~\cite{shu2017study}, with each layer composed of various packages, our result suggests that a smaller number of images in a Chart does not indicate a lower security risk. Practitioners should not solely take the number of images into account while evaluating the Chart security. Our finding aligns with prior work on Docker Hub~\cite{shu2017study} where authors found that the complexity of configurations in Docker images is correlated to the number of vulnerabilities and the vulnerabilities commonly propagate from parent images to child images. 

\label{sec:rq1-results:subsec:cve_distribution}

\begin{Summary}{}{firstsummary}
Out of 13,095 reported Common Vulnerabilities and Exposures (CVEs), 10,982 (83.9\%) are fixable, and most of these CVEs impact more than one Helm Chart. The fixable CVEs predominantly have a high severity rating according to the CVSS V3 specification, indicating they could be easily exploited. Official Charts are equally affected by these fixable CVEs. However, many Charts lack adequate security information for end users. Specifically, 4,833 Charts (representing 44\%) have disabled the feature for security reports, and only 80 (i.e., 1\% of the remaining 56\%) include information about vulnerability fixes in their changelogs. A strong correlation exists between the number of package dependencies in a Chart and its number of fixable CVEs, with a statistical significance (p-value $<$ 0.001).
\end{Summary}

\section{RQ2 results: \RQTwo}
\label{sec:rq3-results}

\textbf{Table~\ref{tab:MitigationStrategies} presents the 11 mitigation strategies belonging to three categories, identified from the 90 Helm Chart GitHub repositories (686 Charts) by open/axial coding iterations until theoretical saturation.} In the iterative axial coding phase, we identified mitigation strategies that belong to three categories: "Ad-hoc", "Automated", and "Informative". We assign the 11 strategies into the categories by observing maintainer interactions with themselves and/or Chart users and evaluating the usage scenarios of each strategy. Next, we present and discuss the empirical observations of the mitigation strategies in a top-down fashion. We first introduce the categories of mitigation strategies assigned via axial coding. Then, we discuss the individual mitigation strategies through our observations, backed by a real-life example on GitHub and a discussion around the maintainer's stance on the strategy. The quoted examples are slightly modified to remove actual usernames and to correct common typos. We present links to all quoted examples as a table in Appendix~\ref{table:appendix_quote}.

\begin{table}[htp]
\scriptsize
\centering
\caption{The 11 CVE mitigation strategies in the 90 associated Chart GitHub repositories (containing 686 Helm Charts)}
\begin{tabular}{lp{6.2cm}r}
\toprule
\textbf{Category} & \textbf{Strategy} & \textbf{\#Repositories} \\
\midrule
\multirow{6}{*}{Ad-hoc}
& [S01] Manual version bump on a CVE report & 31 (34\%) \\
& [S02] Patch application configuration & 13 (14\%) \\
& [S03] Switch to another image variant & 6 (6.\%) \\
& [S04] Customize base OS image & 6 (7\%) \\
& [S05] Minimize functionality for the attack surface & 5 (5\%) \\
& [S06] Enforce the use of image digest & 1 (1\%) \\
\addlinespace
\multirow{2}{*}{Automated}
& [S07] Always release latest dependencies & 29 (32\%) \\
& [S08] Utilize CVE scanners & 3 (3\%) \\
\addlinespace
\multirow{3}{*}{Informative}
& [S09] Define security policy on CVE handling & 22 (24\%) \\
& [S10] Meaningful security changelogs & 11 (12\%) \\
& [S11] Provide CVE mitigation notes & 2 (2\%) \\
\bottomrule
\end{tabular}
\label{tab:MitigationStrategies}
\end{table}
\subsection{Ad-hoc Strategies}

We define an \textit{ad-hoc} strategy as one action requiring significant manual work to prevent, detect and mitigate the impact of CVEs if needed. We mine a total of 6 ad-hoc strategies applied in 36 Chart maintenance repositories.
\hspace{1cm}

\noindent\textbf{[S01] Manual version bump on a CVE report:} This strategy involves a typical step of manually incrementing the dependency (i.e., Chart dependency, image dependency) version. The goal of the strategy is to bump the Helm Chart dependency to a safe version where a known fixable CVE is mitigated by upstream Helm Chart/image maintainers. This strategy is applied to 31 unique Helm Chart repositories. 

\noindent\textbf{Observation:}
This strategy is the most commonly used (31 cases) among all strategies. We observe 2 cases where the initially opened pull request to fix the CVE is insufficient, where the affected dependency (CURL package) appears multiple times in a Chart. Therefore, the fix needs to bump up multiple images to fix the CVEs. Specifically, such a strategy entirely depends on the disclosure date of CVE reports and the availability of maintainers after they become aware of the CVEs and their fixes. In addition, maintainers cannot guarantee the timeliness of CVE mitigation because the CVE reports are created by an end user of the already released Charts in an ad-hoc manner. In 3 cases, maintainers conduct a manual investigation to determine if a CVE truly impacts the Charts before executing the bump-up, while others directly accept the bump-up pull request. [Q01] demonstrates a typical case where a Chart user reported fixable CVEs to the maintainers and asked for a bump-up and rebuild.

\begin{flushright}
[Q01] \textit{Bugs: GHSA-vvpx-j8f3-3w6h (CVE-2022-41723) \& CVE-2022-28391 exists in the current Prometheus image version being used. The bugs have been resolved in Prometheus docker image version v2.43.0. The Prometheus Chart should be rebuilt with this newer docker image...}
\end{flushright}

\noindent\textbf{[S02] Patch application configuration:} This strategy relates to manually patching the vulnerable deployment configurations, once identified, which will prevent potential attacks led by the identified CVE. This strategy is applied in 13 Chart repositories. 

\noindent\textbf{Observation:}
Throughout the analysis, we find that the most frequently (4 cases) patched CVE in this strategy is the Log4Shell vulnerability\footnote{\url{https://nvd.nist.gov/vuln/detail/CVE-2021-44228}}(i.e., CVE-2021-44228). Specifically, we find that this strategy not only appears as a reactive measure to vulnerability reports but also as part of security hardening activities, which requires dedicated knowledge of the Chart dependencies and typically involves multiple rounds of conversation before a patch is merged. [Q02] demonstrates a typical case where a Chart maintainer patched a CVE by specifying a new container configuration.  

\begin{flushright}
[Q02] \textit{Add missing NET\_RAW capability to the system-probe container. This capability is now required with CRIO due to a patch remediation for the CVE-2020-14386...}
\end{flushright}

\noindent\textbf{[S03] Switch to another image variant:} This strategy is an action to reduce potential CVEs. It is applied when a Chart maintainer finds that an existing application image used by the Chart could be replaced entirely by another variant, which is considered less vulnerable to CVEs. This strategy is applied in 6 Chart repositories. 

\noindent\textbf{Observation:}
We find this strategy used in cases where maintainers find an application image variant that frequently contains vulnerable packages that are not actually used by the deployed applications. Such vulnerabilities, although unfixed, are less exploitable given that they cannot be easily triggered by attackers (i.e., not remotely exploitable). The maintainer switches to an equivalent image provided by a third party that is free of the detected CVEs. 

We observe that this strategy receives mixed opinions from Chart maintainers, given that the difference in maintenance practices of the alternative images and default configurations could introduce risks such as incompatibility. [Q03] demonstrates a typical case where the maintainer indicates that switching to an equivalent image is not a good idea but prefers to update the already-used dependency upstream.

\begin{flushright}
[Q03] \textit{Contributor: This PR overrides elasticsearch-exporter tag with latest releases available in docker, whereas main repo is not updated in few months.
Maintainer reply: I don't think it's a good idea to change to the bitnami variant by default, if you want an update, add it upstream.}
\end{flushright}

\noindent\textbf{[S04] Customize base OS image:} This strategy is an action to mitigate OS-level CVEs. The typical step involves migrating a dependency (application) onto a customized OS image that is not affected by the identified CVEs. This strategy is applied in 6 Chart repositories. 

\noindent\textbf{Observation:}
While it is safe in most situations to update and incorporate the CVE fixes from an upstream OS distribution where the CVE patches were submitted~\cite{lin2023vulnerability}, in 2 cases, we find that maintainers manually remove and use alternative packages in order to avoid frequently vulnerable packages like "Curl". This strategy requires building application images "in-house", which demands skills beyond simply switching a pre-built image tag to another variant like S03.

Notably, in 3 cases, contributors suggested rebuilding the application image and switching the underlying OS images to a ``distroless" paradigm. ``Distroless image" is a type of language/application-specific container image that is customized to contain bare minimum dependencies tailored to specific applications~\cite{distroless}. Nonetheless, we find that such ``minimal" images (such as distroless) and their adoption have neither been extensively studied in academia nor widely used in the industry. \citet{haque2022well} conducted a correlation analysis between the number of vulnerabilities and concluded that distroless images, in several cases, suffer from more critical-severity CVEs than their full-size counterpart. The authors suggested that a smaller image size could lead to many severe vulnerabilities. Future research could focus on analyzing the adoption of such images in the Helm Charts and draw empirical understandings of the practical trade-offs of adoption. [Q04] demonstrates an example where the maintainer suggested that adopting distroless images was not always feasible because some applications required various system packages to function.

\begin{flushright}
[Q04] \textit{We are aware of the distroless initiative as well as its pros and cons, in fact, some of the Bitnami containers are distroless. But not always this is an alternative since some applications need different system packages to work so it is needed an underlying distro.}
\end{flushright}

\noindent\textbf{[S05] Minimize functionality for the attack surface:} This strategy is an action that involves a typical step of manually evaluating the deployment definition of Chart containers and their exposed functionalities to external networks such as root access. This strategy is applied in 5 Chart repositories. 

\noindent\textbf{Observation:}
\citet{manadhata2010attack} defined an attack surface as the proportion of resources that attackers use. By doing a manual evaluation, maintainers use their expertise to remove unnecessary permissions and access given to containers and thus reduce the attack surface to vulnerabilities. We find that Chart maintainers accept this strategy and is in alignment with the security best practices on Kubernetes, reported by~\citet{shamim2020xi}. However, we find that reducing attack surface requires extensive knowledge of dependency configuration and mechanisms often maintained by multiple parties. Only in 1 case did the maintainers mention that they would spend time proactively evaluating each Chart to enforce a minimal attack surface. The requirement for significant manual effort and expertise likely contributes to the low occurrence of this strategy in the Chart maintenance repositories. [Q05] demonstrates an example of changing the security context to minimize attack surface for vulnerabilities. 

\begin{flushright}
[Q05] \textit{Reducing kernel capabilities available to a container limits its attack surface containers[] .securityContext .readOnlyRootFilesystem == true}
\end{flushright}

\noindent\textbf{[S06] Enforce the use of image digest:} This is a strategy that is only discussed and executed as a standard in one maintenance repository. However, we observe multiple repositories supplying image digests and tags from the beginning without indicating that it is a CVE mitigation strategy. 

\noindent\textbf{Observation:}
Image digest is a SHA256 hash representation of the layers in a container image. \citet{ohm2020backstabber} reported that unpinned dependencies are among the core reasons for supply chain attacks. In containerization, images could provide multiple digests (builds) even if they share the same image tag. While Charts often point to container images using their tags, industry practice\footnote{\url{https://cloud.google.com/kubernetes-engine/docs/concepts/about-container-images}} suggested that image tags are mutable and could point to unexpected image artifacts upon change. Therefore, Chart users could receive vulnerable images (specified with the same tag) during deployment, even if a CVE check did pass when the Chart was released. Attackers could utilize this flaw in the release process and make critical security consequences. [Q06] demonstrates an action (pull request) to pin a Chart dependency by its digest to ensure a consistent image is used in every deployment.

\begin{flushright}
[Q06] \textit{Using a digest ensures the image being pulled has not been compromised and allows testing release against images without having to tag them.}
\end{flushright}

\subsection{Automated Strategies:}
We define an \textit{automated} strategy as an tool that does not require manual participation beyond the review of pull requests (to approve the actions to execute). Such strategies are inherently automated and can operate independently to prevent, detect and mitigate the impact of CVEs. We mine a total of 2 automated strategies applied to 29 unique Chart maintenance repositories.
\hspace{1cm}

\noindent\textbf{[S07] Always release latest dependencies:} This strategy typically involves automated pipelines that periodically bump up the Chart dependencies. This strategy is applied in 29 Chart repositories. 

\noindent\textbf{Observation:}
This strategy, in a majority of cases, is powered by open-source tools such as Renovate~\cite{renovate}. We find such automated pipelines to run fully automatic in 13 repositories, meaning a commit is submitted to the main branch without maintainer approval. In the remaining cases, we observe that this process requires manual intervention (3 cases), given that a bump-up may yield a defective deployment due to version incompatibility among the dependencies. 

\citet{mirhosseini2017can} observed that repositories with automated pull requests, although benefiting from a quicker merge of updates, often suffer from other downsides, such as accidentally introduced bugs and performance regressions from a newer version. In our study, many automated pull requests mention the versions to update but fail to identify the reasons to update (such as fixing the particular CVEs and the impact of CVEs). In addition, we observe that fully relying on such automated tools does not prevent all fixable CVEs. In fact, we find that among the 20 automated maintenance repositories with a security report, high-to-critical severity (i.e. CVSS score $>$ 7.0) fixable CVEs exist in 18 repositories (90\%). In an extreme case, one Chart artifact is exposed to 713 fixable CVEs, indicating that always releasing the latest dependencies cannot fix all fixable CVEs. We observe such a phenomenon because upstream Chart dependencies could be abandoned or rarely maintained and, therefore, accumulate many CVEs over time.  [Q07] demonstrates an example of automatic bump-up requiring human intervention.

\begin{flushright}
[Q07] \textit{Thanks for letting us know, the new version was detected by our automated test \& release pipeline but we found some issues when packaging this new version and we need to manually fix it. At this moment the team is working on it. We will update this thread once we released the new version.}
\end{flushright}

\noindent\textbf{[S08] Utilize CVE scanners:} This strategy involves an automated CVE scan check before the merge event of pull requests, upon Chart releases and/or on a scheduled timer that is executed in CI/CD pipelines. The goal is to detect CVEs as they are introduced into the Charts. This strategy is applied in 2 Chart repositories.

\noindent\textbf{Observation:} 
We find surprisingly scarce evidence of running CVE scanners in Chart maintenance repositories, although we observe many Chart users mention CVE scanner results in S01. At first glance, it seems counter-intuitive given their efficiency advantages~\cite{brady2020docker,rangnau2020continuous}, and their wide adoption in research studies~\cite{shu2017study,zerouali2022helm}. However, our observation identified two major concerns that account for this phenomenon: the issue of false positives and the need for manual handling, which, in combination, could result in a low adoption rate. 

In issues/pull requests/security policies where the usage of CVE scanners is discussed, we observe that one GitHub repository publishes explicit security guidelines against reporting raw CVE scanning results. We also observe maintainers indicating the problems of false positives (Q08). However, we find it noteworthy that, regardless of the negative stance from the maintainers' perspective, Chart users frequently report CVEs detected by such scanners. Through the iterative open coding phase, we observe CVE reports and fixes proposed based on the results of 7 different scanners, indicating that Chart users relied heavily on such automation before deployment. We observe that two maintainers chose to reject such CVE reports directly and asked the Chart user to propagate the CVE fix request upstream, even when both the Chart source code and the image dependency are maintained under the same organization (official Charts). Quoted example [Q08] shows a typical example of false positives via CVE scanners, where the mentioned Log4Shell CVE has been addressed by patching application configuration (S02).
\begin{flushright}
[Q08] \textit{Thanks for a quick response on this, I guess I missed out on this release note, appreciate sharing this. Also we are looking forward for 7.16.2 to mitigate the false positive too from scan results.}
\end{flushright}

\subsection{Informative Strategies:}
We define an \textit{informative} strategy to inform practitioners of the occurrence of CVEs and their mitigation plans. The informative strategies contain (1) the instructions on how to mitigate a CVE for practitioners and (2) the channels where the patch of the CVE was discussed and its report was submitted. We mine a total of 3 informative strategies applied to 27 unique Chart maintenance repositories.
\hspace{1cm}

\noindent\textbf{[S09] Define security policy on CVE handling:} This strategy involves presenting information about reporting, disclosing and finding information on user-impacting CVEs. This strategy is applied in 22 Chart repositories.

\noindent\textbf{Observation:} Although 22 out of 90 GitHub repositories contain a ``SECURITY.md," which is an automatically parsed file to populate the "Security policies" tab of GitHub, we observe only two security policies providing practical information relevant to Helm Charts. Specifically, among the 22 repositories employing such a security policy, 20 (90\%) directly reference an organization-wide policy not tailored to the Helm Charts. We find such security policies to be generally less informative and to focus solely on the core application dependency of the Chart (owned by the organization). Example Q10 demonstrates an informative security policy on "Open CVEs" (i.e., fixable CVEs) in the Bitnami organization, explaining why the CVEs remained unfixed in the Charts because these CVEs are indeed unfixable.

\begin{flushright}
[Q09] \textit{Open CVEs are the ones that have not been fixed by the Linux Distribution maintainers because they did not work on that yet or they do not consider a critical issue. Bitnami is not able to fix those CVEs since those fixes depend directly on the distribution maintainers.}

\textit{Is Bitnami Releasing Charts Or Containers That Include CVEs?} 

\textit{All Bitnami containers and Helm Charts do not include fixable CVEs.}
\end{flushright}

\noindent\textbf{[S10] Meaningful security changelogs:} This strategy involves presenting information around the security changes such as CVE-IDs, impact and suggestions for update. This strategy is applied in 11 Chart repositories.

\noindent\textbf{Observation:} \citet{abebe2016empirical} identified release notes as an important source of software changes that could guide software users through caveats. In most cases, we find that Chart maintenance repositories do not keep meaningful changelogs that indicate the changes made between two consecutive Chart releases. We find that such a phenomenon is mainly attributed to two situations: (1) unawareness of tools (S07) that could help formulate changelogs, and (2) the frequent releasing nature of Charts implies significant manual effort to keep track of all changes. 

Although Artifact Hub defines a special annotation to record the changelogs: "annotations: artifacthub.io/changes''\footnote{\url{https://artifacthub.io/docs/topics/annotations/helm/}}, we observe only 5 out of the 11 GitHub repositories use such annotations to disclose CVEs. In the case of missing meaningful changelogs, as there are no simple methods (other than tracing all commits and pull requests in the repository) to understand the actual changes, users could be forced to use the Charts without knowing the existence and/or impact of vulnerabilities. [Q10] shows a practical usage of the changelog annotation to inform Chart users of specific changes in a new Chart version.

\lstdefinestyle{yaml}{
     basicstyle=\color{black}\footnotesize,
     rulecolor=\color{black},
     string=[s]{'}{'},
     stringstyle=\color{black},
     comment=[l]{:},
     commentstyle=\color{black},
     morecomment=[l]{-}
 }

\begin{flushright}
\begin{lstlisting}[style=yaml]
[Q10]
artifacthub.io/changes: |
- kind: security
  description: 'loki-stack: update Chart from 2.9.9 to 2.9.10'
  links:
    - name: 'fix: update bats image for CVE-2019-14697'
      url: Commit Hash (Omitted due to length)
\end{lstlisting}
\end{flushright}

\noindent\textbf{[S11] Provide CVE mitigation notes:} This strategy provides detailed mitigation methods for particular CVEs discovered in a Chart. This strategy is applied in 2 Chart repositories.

\begin{figure}[htp]
    \centering
    \includegraphics[width=\textwidth]{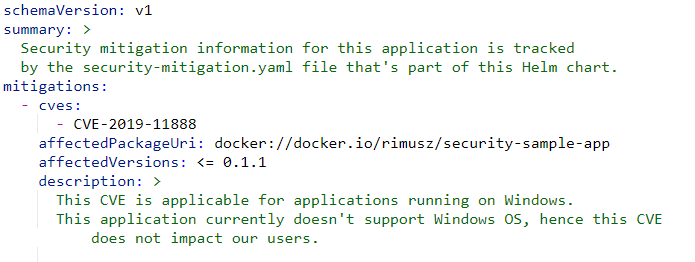}
    \caption{A sample CVE mitigation note from JFrog, showing the best practice to disclose security changes to Chart users.}
    \label{fig:specification}
\end{figure}

\noindent\textbf{Observation:} We find this method to be applied rarely. In two cases, it is used for the well-known Log4Shell vulnerability. One of the now-archived alternatives (JFrog ChartCenter) to Artifact Hub encourages such a strategy by defining a specification named "Security Mitigation Notes''\footnote{\url{https://jfrog.com/blog/helm-chart-security-mitigation-in-chartcenter/}}. Figure~\ref{fig:specification} depicts an example of implementing the specification through a file that ships along with the Helm Chart artifact. The specification involves manual notes for mitigation strategies and the impact radius of a CVE (version range and package). 

Deploying such a strategy may give users confidence when deploying a third-party Chart since users could follow the notes to upgrade Chart versions to a vulnerability-free release. Writing the mitigation notes could be tedious for Chart maintainers in practice. We find that such CVE mitigation notes could complement strategy S10 by clarifying a detailed description of CVE exploitability and impact in the context of the affected Chart (i.e., adding CVSS temporal and environmental information~\cite{ali2011software}). Future studies could explore the historical adoption rate of CVE mitigation notes and propose automation to generate such notes using information from the National Vulnerability Database (NVD) and/or through static/dynamic analysis of the deployed resources/configuration by a Helm Chart. [Q12] demonstrates a quote that explains the reasoning behind the "Security Mitigation Notes" specification.

\begin{flushright}
[Q11] \textit{This enables Helm Chart authors to annotate their Charts with notes about the CVEs flagged by their dependencies, letting users know whether and when to be concerned, or if they can mitigate the risk. It helps you to say in response to a CVE, "Yes, but…" and engage in a type of dialogue with the users of your Chart.}
\end{flushright}

\subsection{Cross-referencing the Strategies: Dependency Bump-up is Not All We Need}

\textbf{Simple bump-up for Chart dependencies is common (S01, S07) to ease the manual work of following upstream releases. Nevertheless, such a strategy alone does not secure Helm Charts, even with automation.}
Previous studies stress the importance of adopting automation in the domain of DevSecOps~\cite{tomas2019empirical,rangnau2020continuous,kreitz2019security}. Yet, the authors also reported that existing automation is little known to practitioners due to a lack of security-related knowledge and tool standards. Such observation aligns with the phenomenon in the Helm Charts domain, as 31 repositories bump up versions manually, suggesting that maintainers may not be aware of automation that could assist in maintaining up-to-date dependencies in Charts. Overall, we observe that fewer than 50\% of the studied Chart maintenance repositories adopt the S07 strategy.  

Nevertheless, our analysis for S07 reveals that high and critical severity CVEs remain unfixed (despite being fixable) in those Charts that always use the latest dependency versions (i.e., within 18 repositories). In most cases, the Chart maintainers do not directly build the dependencies but wait (with/without automation) for potential upstream mitigation to be available. We also observe that automated version bump-ups led to incompatibility issues due to dependencies that could not work with a newer version, which leads to maintainer overhead~\cite{rombaut2023there}. This phenomenon indicates that trivial automated fixes cannot fully mitigate fixable vulnerabilities. 

\textbf{Informational and ad-hoc strategies serve as a necessary complement to automated strategies to secure Helm Charts.} We find that CVE scanners (S08) are often prone to false positives and thus may only accurately represent the actual security of Charts if maintainers provide additional information that is tailored to the Chart (S10, S11). Mitigation notes and changelogs with additional information could convey effective CVE fix information to users and help assess vulnerability impacts. On the other hand, manual work, such as switching to another image variant (S03) and customizing base OS images (S04), could proactively mitigate future CVEs originating from the vulnerable images built by upstream maintainers.

\subsection{Formulating the Grounded Theory: Responsibility and Trade-off}
Figure~\ref{fig:grounded} presents a detailed view of our final grounded theory. With the 11 CVE mitigation strategies and categorizing strategies into three distinct categories (i.e., Ad-hoc, Automated and Informative), we conduct selective coding to draw commonalities from the strategies before formulating the final theory. 

The core subject from the studied data is "CVE mitigation strategies in Helm Chart maintenance repositories." We identify two driving factors that fundamentally relate to the high prevalence of fixable CVEs that remain unfixed: Responsibility and Trade-off, which emerged as central keywords interconnecting the observed maintainer stances on CVE mitigation strategies. We formulate the grounded theory as follows: \textbf{The high number of fixable CVEs that remain unfixed in Helm Charts arises from unclear responsibilities and trade-offs between stability and security. Shared responsibility guidelines are needed to secure the Helm Charts better.}

\subsubsection{Shared CVE mitigation responsibilities}
\textbf{There do not exist precise practices or guidelines related to who and where vulnerabilities should be mitigated.} 
The statement "\textit{I don't think it's a good idea to change to the bitnami variant by default, if you want an update, add it upstream.}" from a Chart maintainer in [Q03] indicates that the Chart maintainer was not responsible for addressing the fixable CVE and asked to coordinate with the upstream dependency. While there exists a common practice that bugs should be fixed by the upstream projects in the Linux distributions\cite{lin2022upstream}, the Helm Chart ecosystem may not have such existing guidelines. 

Adopting a shared-responsibility model would be a way to mitigate fixable vulnerabilities. Previous studies on DevSecOps~\cite{mao2020preliminary,myrbakken2017devsecops} suggested that the principle of shared-responsibility is attracting increasing attention from academia and the industry. The shared-responsibility principle describes preemptively mitigating security concerns, such as CVEs, as early as possible in the software supply chain. Implementing the principle requires a collective approach for Helm Chart stakeholders (i.e., image/package dependency maintainers, Chart maintainers) to minimize security risks such as fixable CVEs. \citet{zahan2022weak} observed that attackers often aim at the weakest package dependencies to attack, which effectively emphasizes a need to secure all dependencies in a Chart. Such delays suggest the necessity of applying the shared-responsibility model~\cite{mao2020preliminary,tomas2019empirical} in the Helm Chart domain, where all stakeholders, including Chart maintainers, should proactively participate in securing the Charts.

\textbf{Chart maintainers lack incentives to motivate mitigation actions.}
Although it is possible to maintain and release Helm Charts that are unaffected by fixable CVEs (Q10), we did not find evidence where the Chart maintainers have sufficient incentives to enforce such a release practice and/or share the maintenance responsibility. 

Currently, to mitigate fixable CVEs that remain unfixed in dependencies, maintainers could rebuild and/or swap out outdated packages for their safer equivalents (S03, S04). However, maintainers are typically reluctant to do so due to the additional manual efforts\footnote{\url{https://github.com/kubernetes/ingress-nginx/issues/8520}}. A prior work~\cite{kula2018developers} reported that software developers are usually unaware of security practices in software maintenance and often consider security-hardening activities an extra overhead. In addition, 
\citet{liu2020understanding} reported that CVE mitigation is frequently delayed in Docker images due to a lack of incentive and the complex composition nature of images. In the context of Helm Chart maintenance, we observe that the Chart maintainers suggested users report CVEs upstream, i.e., to the maintenance repository of dependencies (package and Chart dependencies). A Chart maintainer in the following example [Q13] stated a strong message asking for the delegation of CVE mitigation back to the upstream dependencies. 

\begin{flushright}
[Q13]    \textit{It doesn't make any sense to provide these reports for helm Chart. Please forward this to identified projects.}
\end{flushright}

Mitigating CVEs in upstream dependencies would be an ideal way to reduce security risks. However, Helm Charts involve nested structures, as discussed in Section \ref{subsec:background-charts-dependencies}, each with unique maintenance practices and security policies. As upstream dependencies may not release a fix promptly, Chart users would expect the Chart maintainers to be the delegation of finding CVE mitigations and applying them. As a result, delegating mitigating CVEs to Chart maintainers results in a dilemma of fixable CVEs that remain unfixed in the Helm Charts (the result in RQ1). Therefore, current Chart users must identify, understand and mitigate the fixable CVEs themselves.

\subsubsection{Trade-offs when mitigating CVEs in Helm Charts}
\textbf{The adoption of CVE mitigation strategies is subject to trade-offs. Maintainers may prioritize the maintenance effort and/or the stability of Helm Charts over security.}
Among all three categories of CVE mitigation strategies, we observe that none could eliminate all the fixable CVEs. In addition, we observe that the 11 CVE mitigation strategies receive a mixture of maintainer stances, given the trade-offs between security, resources (e.g., time), and non-functional requirements such as stability and compatibility~\cite{chung2012non}. For example, a Helm Chart could introduce backward incompatible features (e.g., refactored APIs) when bumping up an image dependency (to incorporate a CVE fix). We consider such trade-offs crucial in the context of Helm Charts since Helm deployments often scale across large distributed systems and attract adoption concerns around usability and security~\cite{li2019service,chen2023practitioners}. 

\textit{Stability vs. security:} In two strategies [S03, S04], we observe that the maintainers prioritize the stability of Docker images over security benefits, suggesting the need to ensure the successful operation in an application.~\citet{zerouali2021multi} found that Docker images are often outdated regarding vulnerability and bugs, whereas stable OS images are consistently more vulnerable. As a result, it is understandable that current Charts could be exposed to fixable CVEs that are only patched in a non-stable base image, which Chart maintainers are reluctant to accept. We consider this preference for stability over their potential vulnerability a paradoxical scenario where the very act of ensuring application functionality (via stable images) may expose the application to security risks due to latent vulnerabilities in these dependencies. 

Example Q14 demonstrates a statement around such trade-offs between security and stability from the Chart users' and Chart maintainers' points of view. 

\begin{flushright}
[Q14]   \textit{User: Therefore, it would be great if we could do a new 2.8.27 hotfix which simply has no functional changes but pulls in Debian updates.}

\textit{Maintainer: I am sorry, but we only release containers for the stable version of the asset.}
\end{flushright}

\textit{Maintenance effort vs security:} During the open coding phase, we find that ad-hoc configuration patching (S02) and manual inspections to minimize functionalities (S05) cannot be replaced by any existing automated strategy. The inherent complexity and rapidly changing nature of Helm Charts and their dependencies increase the difficulty of Chart maintenance. We consider that such a trade-off originates from a shortage of resources and/or expertise within maintenance teams to identify and mitigate vulnerabilities~\cite{tomas2019empirical} proactively. This trade-off is also magnified by the unclear boundary of responsibilities discussed in the previous section. If these strategies were more effectively employed (i.e., through security guidelines tailored for Helm Charts and automated assistance), they could significantly reduce the prevalence and impact of unfixed high-severity and critical-severity CVEs.

\begin{Summary}{}{secondsummary}
We present a comprehensive discussion of 11 identified CVE mitigation strategies grouped into three categories in 90 GitHub Chart maintenance repositories associated with 686 Charts. We distill our findings into a GT where responsibility and trade-off emerge as the key factors influencing adopting the identified strategies. 6 of the 11 strategies are ad-hoc, 2 are automated, and 3 are informative. Due to the high number of dependencies in a Chart, Chart maintainers often need to manually investigate how the upstream dependency mitigates a vulnerability before incorporating the fix. The automated tools have limitations (e.g., false positives in CVE scanners) that require Chart maintainers to take additional effort while mitigating vulnerabilities. The information provided by Chart maintainers needs more precise and clear instructions on how to mitigate vulnerabilities for Chart users.
\end{Summary}

\section{Implications} 
\label{sec:implications}
\subsection{Implications For Researchers and Helm Chart Maintainers} 
\label{sec:implications:subsec:researchers}

\textbf{Researchers and Chart maintainers could investigate guidelines related to the shared-responsibility model for Chart maintenance.} 
The results in RQ1 indicate that a small set of 20 vulnerabilities affect at least 1,300 Helm Charts. The maintainers of such a large number of Helm Charts could coordinate with each other to share the responsibility of mitigating the 20 vulnerabilities. In such a way, the maintainers of each Helm Chart would reduce the individual effort, and the mitigation steps could be shared across Charts. Our findings in RQ2 suggest that Chart maintainers assume mitigating vulnerabilities as a low priority and out of their responsibility. A guideline for sharing vulnerability information across Charts would be a first step. For example, maintainers could build a list of Helm Charts affected by a common fixable vulnerability, its available fix and whether it has been incorporated in a particular Chart, so maintainers could either refer to the Chart or link to the available fix. As Helm Charts are one of the containerization technologies, researchers could refer to security practices in the more mature technologies (e.g., Docker) and take into account the nature of Helm Charts (e.g., the complexity of a Chart, lack of security guidelines) while developing security practices of the shared-responsibility model for Helm Charts.

\textbf{Researchers and Chart maintainers could develop efficient and trust-worthy automation for securing Helm Charts}. 
RQ1 results indicate that the number of fixable CVEs in Charts significantly correlates with the number of package dependencies instead of image dependencies. Official Charts are not inherently more secure than unofficial Charts. Our finding suggests that current Chart users may need more effective means to understand the security of Helm Charts by such metadata. The findings in RQ2 indicate that the automated strategies only account for 18\% of the identified strategies (i.e., 2 out of 11 strategies) because of the shortcomings of the automated strategies (e.g., false positives). In addition, the findings in RQ2 highlight complex trade-offs, i.e., stability vs. security and additional maintenance effort vs. security. The complexity of security threats continues to evolve, and the need for scalability becomes pressing in cloud-native technologies, such as service mesh~\cite{li2019service}. As such, developing reliable automation would increase the use of automated strategies, reducing the security risks in the Helm Charts ecosystem and erasing the unwillingness of maintainers to adopt such automation. Given the complex dependencies in a Chart, researchers could investigate why the existing CVE scanners report many false positives and guide maintainers on building an efficient and trustworthy automation tool.

Furthermore, future researchers and Chart maintainers could track the results (e.g., logs, discussions) of using existing tools to improve the adoption of automated CVE mitigation strategies. For example, the tool may take much time to complete a scanning task of a Chart. Researchers could develop a better algorithm to reduce the time taken. Maintainers could build tools that automatically generate mitigation notes and changelogs. We also suggest studying the impact of continuously vulnerable Charts to gain insights into whether it is wise to sacrifice security through trade-offs.

\textbf{Researchers could coordinate with Chart maintainers to develop a mechanism to notify users of precise security-related information (e.g., fixed vulnerabilities and mitigation plans of fixable vulnerabilities)}.
Our findings in RQ1 indicate the low utilization of security features, such as security advisory and changelogs (see Table \ref{table:informative-metrics-github}). Given the large number of false positives generated by CVE scanners (as discussed in [S08]), Chart maintainers and Chart users cannot rely on CVE scanners to identify vulnerabilities in a Chart.
As Chart users may not have sufficient expertise in the applications deployed by Helm Charts, it becomes difficult to accurately assess the impact of detected CVEs without information from the experts (i.e., Chart maintainers). We thus recommend that future researchers and Chart maintainers explore ways to dynamically locate vulnerabilities with actual impacts based on the dependency and configurations of a Chart. 

In addition, meaningful security changelogs (S11) and security mitigation notes (S12) are noticeably absent in Helm Chart releases (i.e. cross referencing RQ1 and RQ2 results). Such information would ease security concerns related to Chart vulnerabilities. As the release schedules of package and Chart dependencies in a Chart are dynamic, we suggest future researchers investigate the potential automation of CVE analysis and mitigation. This would not require manual effort from maintainers and may leverage large language models (LLMs) that are dedicated to evaluating Kubernetes manifests to do so~\cite{chen2021evaluating,k8sgpt}. Future researchers could build a predictive model for identifying potential vulnerabilities in Helm Charts and suggest mitigation methods using the CVE database and available mitigation strategies.

\textbf{Researchers and Chart maintainers could use the 11 identified mitigation strategies (RQ2) as a fundamental checklist to enhance Chart security}. 
In RQ2, for each strategy, we explain why maintainers employ such a strategy and the shortcomings of using such a strategy. Maintainers could learn from the strategies and select suitable ones based on their constraints (e.g., resources), thus balancing the maintenance effort and security risks.
Researchers can use the checklist to boost the security of Charts by analyzing the takeaway from each mitigation strategy, looking for empirical evidence, testing the mitigation strategies, and evaluating their effectiveness on a large scale. 

On the other hand, our study underscores that automated strategies play an essential role in CVE mitigation. However, they must be complemented with manual and informative strategies to address vulnerabilities effectively. When implementing these mitigation strategies in Helm Chart maintenance, current Chart maintainers could use the checklist to balance the trade-offs among security and stability, resource constraints, and effort.

\subsection{Implications For Chart Users}
\label{sec:implications:subsec:users}

\textbf{Chart users should be aware that Charts are affected by a median of 119 unfixed vulnerabilities and conduct a comprehensive security assessment before deploying Charts into their systems}. The results in RQ1 suggest that the number of vulnerabilities is correlated with the number of dependent packages in a Helm Chart instead of the number of images. In addition, official Helm Charts are also affected by fixable vulnerabilities, contrary to the finding in the prior work in Docker~\cite{zerouali2019impact}. Furthermore, in RQ2, we observe that Helm Charts are still affected by high and critical severity vulnerabilities, even when applying automated strategies. We suggest Chart users adopt vulnerability scanners to uncover vulnerabilities not disclosed by maintainers and promptly evaluate the security risks of the systems before deploying public Helm Charts. 

\section{Threats to Validity}
\label{sec:threats}

Below, we discuss threats to the study validity and the methodologies we applied to minimize these threats, based on literature guidelines~\cite{wohlin2012experimentation}.

\textbf{Internal Validity:} In RQ2, we utilized a GT methodology, which can provide useful insights if applied correctly despite being inherently qualitative and subjective. To address the threat, we followed the guidelines and protocols widely used by prior work~\cite{zimmermann2023grounded,draucker2007theoretical,hoda2012developing,carver2007use}. We conducted multiple (10) iterations of the GT coding phases, and involve three researchers (i.e. first and second authors iteratively conduct and discuss during coding phases, subsequently discuss and confirm insights with the third author) to discuss our observations before arriving at the generalized theories, which minimized the risk of interpretation bias. Given the highly technical nature of the CVE mitigation strategies, our grounded theory is less susceptible to biases, as we support every strategy we discuss with relevant quotes and literature references. Appendix~\ref{table:appendix_quote} includes all quoted examples and their URLs for future verification. 
In addition, during the theory-building phase, we cross-referenced our findings with previous work to ensure our observations were well-founded and robust.

In the qualitative analysis, it is possible that one could generalize CVE mitigation strategies based on different aspects. In RQ2, we took several measures to reduce this threat. We iteratively conducted axial coding phases to ensure that the categorical labels faithfully and accurately represent the various CVE strategies. Although alternative aggregation of the mitigation strategies could exist, we believe our current categorization is meaningful to draw practical takeaways.

We acknowledge the possibility of an alternative path of evaluating mitigation strategies through analyzing individual CVEs and tracing their mitigation in each development repository, which is used by one of the recent studies of vulnerabilities in Linux distributions~\cite{lin2023vulnerability}. While such a methodology is effective for studying the mitigation strategies when the studied subjects are well-constrained (e.g., Debian vs Fedora), we observe the Helm domain having unique characteristics given the large number of Charts with security reports (6,202) and unique fixable CVEs (10,982), as introduced in Section~\ref{sec:methodology}. Thus, we follow previous literature to conduct theoretical sampling and reach theoretical saturation~\cite{draucker2007theoretical}. We generalize mitigation strategies into a grounded theory that could help explain the prevalent existence of fixable CVEs that remain unfixed.

\textbf{External Validity:} We acknowledge that some CVE  reports and their corresponding fixes might be temporarily kept private due to the embargo policy~\cite{ramsauer2020sound, lin2023vulnerability}. However, as a grounded theory research, we have comprehensively investigated 90 GitHub Chart maintenance repositories (686 Charts) and reached theoretical saturation. Through theoretical saturation, we ensured that we did not encounter any new CVE mitigation strategies when exploring new repositories beyond our dataset. We believe that our analysis provides a representative understanding of Chart maintenance activities and mitigation strategies that are employed and discussed by maintainers. 

Another potential threat to external validity is the reliance on Artifact Hub data for Helm Charts, which may not represent the complete landscape of public Helm Charts. Charts may be hosted on other platforms (e.g., JFrog Chart Center) or private registries. However, Artifact Hub is a popular repository where more than 11,035 Charts are available for users (as mentioned in Section \ref{sec:background}) and the other platforms, such as JFrog has recommended migrating to Artifact Hub\footnote{\url{https://jfrog.com/blog/into-the-sunset-bintray-jcenter-gocenter-and-chartcenter/}}. Therefore, we believe our study, while not exhaustive, presents a representative sets of vulnerabilities in Helm Charts and their mitigation strategies.
\section{Conclusions}
\label{sec:conclusions}

This paper presents a mixed-methods empirical study on a set of 11,035 Charts on the Artifact Hub registry affected by at least 13,095 unique vulnerabilities. Our work systematically investigates the vulnerabilities through a combination of quantitative analysis and qualitative analysis. Our study has verified that fixable CVEs that remain unfixed are dominant across Helm Charts and has provided in-depth discussions regarding the employed mitigation strategies, with a curated theory that revolves around two  concepts, including responsibility and trade-off, to explain the phenomenon.

We conduct statistical tests to understand the prevalence of CVEs in Helm Charts and provide insights based on the complexity of a Helm Chart, the CVSS metrics, and the utilization of security features.
Our results shed light on the studied Charts, which are vulnerable to high-severity vulnerabilities that are easy to exploit. A larger number of package dependencies correlates with a large number of vulnerabilities.
Lastly, using grounded theory to study mitigation strategies employed by Chart maintainers, we uncover 11 mitigation strategies in three categories (i.e., ad-hoc, automated, and informative) from the GitHub repositories associated with Helm Charts. Chart maintainers often use ad-hoc strategies to mitigate vulnerabilities as automated tools have side effects, such as false positives, that adds additional burdens to maintainers.

Given a large number of unfixed vulnerabilities (i.e., 10,982 fixable and 2,113 not-patch-yet ones) that have a median of high severity, we recommend that researchers and Helm Chart maintainers investigate guidelines for adopting a share-responsibility model to mitigate vulnerabilities and develop more reliable and accurate automated tools for preventing security threats.
Future work could enhance the adoption rate and effectiveness of automated CVE mitigation strategies and encourage the usage of informative security features. We advise Chart users to conduct comprehensive security assessments before deploying Charts into their systems. These combined efforts could improve the security of Helm Charts and the deployed distributed systems.

\section{Conflict of Interest}
All authors declare that they have no conflicts of interests.

\section{Data Availability Statement}
The datasets generated and analyzed during the current study are available from the corresponding author upon reasonable request.

\appendix
\section{Appendix}
\begin{table}[!ht]
\scriptsize
\caption{Representative examples and URLs from GitHub Chart maintenance repositories}
\begin{tabular}{l p{10.6cm}}  
\hline
\textbf{ID} & \textbf{URL} \\ \hline
 Q01 & https://github.com/prometheus-community/helm-charts/issues/3139 \\\hline
 Q02 & https://github.com/DataDog/datadog-operator/pull/246 \\\hline
 Q03 & https://github.com/prometheus-community/helm-charts/pull/2957\\\hline
 Q04 & https://github.com/bitnami/charts/issues/13516\\\hline
 Q05 & https://github.com/hazelcast/charts/issues/21\\\hline
 Q06 & https://github.com/helm/charts/issues/13449\\\hline
 Q07 & https://github.com/bitnami/charts/issues/17382\\\hline
 Q08 & https://github.com/bitnami/charts/issues/8449\\\hline
 Q09 & https://github.com/teutonet/teutonet-helm-charts/\\\hline
 Q10 & https://docs.bitnami.com/kubernetes/open-cve-policy/\\\hline
 Q11 & https://github.com/adfinis/helm-charts\\\hline
 Q12 & https://jfrog.com/blog/helm-chart-security-mitigation-in-chartcenter/\\\hline
 Q13 & https://github.com/prometheus-community/helm-charts/issues/2582\\\hline
 Q14 & https://github.com/bitnami/charts/issues/12706 \\\hline
\end{tabular}
\label{table:appendix_quote}
\end{table}

\end{document}